\documentclass[a4paper, amsfonts, amssymb, amsmath, reprint, showkeys, nofootinbib, twoside, floatfix]{revtex4-1}
\usepackage[english]{babel}
\usepackage[utf8]{inputenc}
\usepackage[colorinlistoftodos, color=green!40, prependcaption]{todonotes}
\usepackage{amsmath}
\usepackage{float}
\usepackage{adjustbox}
\usepackage{placeins}
\usepackage{bm}
\usepackage{xcolor}
\usepackage{multirow}
\usepackage{textgreek}
\usepackage{amsthm}
\usepackage{mathtools}
\usepackage{physics}
\usepackage{xcolor}
\usepackage{graphicx}
\usepackage[left=23mm,right=13mm,top=35mm,columnsep=15pt]{geometry} 
\usepackage{adjustbox}
\usepackage{placeins}
\usepackage[T1]{fontenc}
\usepackage{lipsum}
\usepackage{csquotes}
\usepackage[pdftex,
    colorlinks=true,
    linkcolor=black,
    citecolor=black,
    urlcolor=blue,
    pdftitle={Article},
    pdfauthor={Author}
]{hyperref} % For hyperlinks in the PDF
\begin{document}
\title{Nano-Clay-Stabilized Water-in-Oil Colloidal Pickering Emulsions as Thixotropic Lubricant}

\author{Arun Kumar,\textit{$^{a}$}  Rahul Yadav,\textit{$^{b}$} Yogesh M. Joshi,\textit{$^{b*}$} Manjesh K. Singh\textit{$^{a}$}}
    \email[Correspondence email address: ]{joshi@iitk.ac.in; manjesh@iitk.ac.in}
%    \thanks{$\dag$ Equal contribution}
    \affiliation{$^{a}$~Department of Mechanical Engineering, Indian Institute of Technology Kanpur, Kanpur-208016, Uttar Pradesh, India\\
$^{b}$~Department of Chemical Engineering, Indian Institute of Technology Kanpur, Kanpur-208016, Uttar Pradesh, India}

% \date{\today} % Leave empty to omit a date

\begin{abstract}
The limitations of conventional mineral oil–based lubricants motivate the development of environmentally benign emulsions capable of providing lubrication and heat dissipation in demanding applications. In this study, nano-organoclay (Garamite 1958\textsuperscript{\textregistered})-stabilized thixotropic water-in-oil Pickering emulsions are developed using sunflower oil as the base. The rheological and tribological properties of the emulsion system are systematically examined. Rheological findings reveal a pronounced increase in yield stress, shear thinning and thixotropic behavior on increasing Garamite loading percentage in the emulsion.  The tribological performance is assessed against dry, water, and oil-lubricated conditions for a steel–steel interface under high contact pressure. The findings indicate that the tribological performance is significantly influenced by the microstructure and thixotropic behavior of the emulsions. The emulsion with the optimal nano-clay concentration demonstrates approximately 41\% and 84\% lower friction and approximately 80\% and 96\% lower wear than oil and water, respectively. The emulsion exhibits sensitivity to the sliding direction and displays load-responsive friction behavior with a memory effect owing to the reversible structuring of the clay–droplet network. This superior performance is attributed to the combined effects of thixotropy, anisotropic nanoclay morphology, and stable droplet armoring, which form a robust and adaptive interfacial film. This study advances the understanding of Pickering emulsions in metallic tribosystems by correlating the microstructure and rheology with tribological performance, thereby facilitating the design of high-performance, smart, and eco-conscious lubricants for metallic systems.
\end{abstract}


\maketitle
\section{Introduction}\label{sec1}

Conventional mineral oil-based lubricants raise environmental concerns \cite{Singh2022AqueousLubrication}. Water is environmentally friendly, inexpensive and has low shear resistance and high thermal conductivity. However, its poor pressure–viscosity response limits its load-bearing capacity and restricts its direct use as a lubricant \cite{Singh2015PolymerExperiments}. Therefore, emulsions that function as both lubricants and coolants are widely used in metal processing and heavy-duty engineering applications \cite{Chen2022Surfactant-ModifiedAnticorrosion, Bao2022AEmulsion}.

 Pickering emulsions offer advantages over conventional surfactant-stabilized emulsions, such as low toxicity, higher stability, cost-effectiveness, and ease of recovery \cite{Liu2024TribologicalSurfactants}. Although Pickering emulsions are extensively used in food \cite{Du2025InvestigationAnalogues,Feng2025Cinnamaldehyde-loadedPerspectives,Du2025WhippedCrystals,Lu2021AParticles}, biomedical \cite{Guo2025Metalloparticle-EngineeredPathogens,Dai2023InductionGel,Xia2025AChemoembolization,Zhang2025UntyingBio-Media,Guan2026EngineeringCarriers,Yao2025NaturalBacteria} and 3D/4D printing \cite{Yu2024VersatileAssemblies,Wu2021AttractiveGels,Zhang2024JammedGels}, their tribological applications, particularly under hard contact conditions \cite{Chen2022Surfactant-ModifiedAnticorrosion,Liu2024TribologicalSurfactants,Wu2024NaturalLubrication,Yu2024VersatileAssemblies,Zhang20254D-PrintableSurfactants,Wu2017TheEmulsion,Yang2019Branch-chainEmulsions,Xu20252DApplication} remain limited. In tribology, Pickering emulsions are primarily employed either as templates for fabricating antiwear composites or directly as lubricants \cite{Bao2022AEmulsion}. Their superior performance arises from mechanisms such as repairing effects \cite{Liu2024TribologicalSurfactants}, strong tribofilm formation, ball-bearing action \cite{Yu2024VersatileAssemblies} and interlayer sliding in layered particles \cite{Xu20252DApplication}. In addition to these well-recognized mechanisms, the role of thixotropy in enhancing the lubrication performance of aqueous thixotropic gels has been highlighted in our previous studies \cite{Kumar2024TribologicalNanoparticles,Kumar2025Nano-silicaLubricant}. In thixotropic systems, the viscosity decreases owing to microstructural breakage under the influence of a deformation field (shear flow). Upon cessation of the flow, the viscosity recovers owing to microstructure buildup over time. Such rapid microstructural adaptation to changing shear conditions is advantageous for lubrication under varying conditions of load, speed, and direction of motion.

 Pickering emulsion stabilizers range from natural to synthetic particles with diverse morphologies. They include graphene oxide \cite{Yu2024VersatileAssemblies,Wu2017TheEmulsion}, silica particles \cite{Chen2022Surfactant-ModifiedAnticorrosion,Guan2026EngineeringCarriers,Yao2025NaturalBacteria}, metal–organic framework nanoparticles \cite{Li2025AdvancedEvaluation}, ZnO particles \cite{Liu2024TribologicalSurfactants,Cionti2022One-stepFunctionalization}, magnetic particles \cite{Chen2022Surfactant-ModifiedAnticorrosion,Liu2024TribologicalSurfactants}, polymer \cite{Yu2024VersatileAssemblies,Silmore2016TunablePGLNs}, proteins \cite{You2023TribologyMicrogels}, cellulose \cite{Bertsch2025PickeringEnhancers}, quantum dots \cite{Huang2025Palladium/grapheneBenzylamine}, Janus particle/nanotubes \cite{Zhao2025Dumbbell-shapedEmulsions} and clay nanoparticles \cite{Liu2024TribologicalSurfactants,Stehl2020Oil-in-WaterFilterability,Merad2021RheologicalClay,AssuncaoDorigon2025DevelopmentNanoclay,Kang2024AActivity}. %Recently, amphiphilic nanoparticles have attracted attention owing to their enhanced interfacial activity and tribological performance \cite{Bao2022AEmulsion}. Notably, nanoparticle-stabilized emulsions exhibit exceptional stability against temperature, pH, and ionic variations, making them promising candidates for advanced lubrication systems \cite{Wu2017TheEmulsion}. 
Nanoclays, as renewable and sustainable materials, have gained significant attention. Their high surface activity and strong interfacial affinity enable efficient adsorption at oil–water interfaces, making them effective stabilizers for Pickering emulsions \cite{Lu2021RecentApplications,Lisuzzo2022PickeringApplications}. These emulsions have been applied in water-based drilling fluids \cite{AssuncaoDorigon2025DevelopmentNanoclay,Liu2021PickeringConditions}, antibacterial systems \cite{Kang2024AActivity}, and fracturing fluids \cite{Luo2025RheologicalFluids}. Their diverse anisotropic morphologies (nanorods, fibers, tubes and sheets) further allow precise control over the interfacial structure and rheological behavior \cite{Lisuzzo2022PickeringApplications}. Building on this, Dorigon et al. \cite{AssuncaoDorigon2025DevelopmentNanoclay} demonstrated that bentonite nanoclay, primarily composed of platelet-shaped montmorillonite, forms highly stable Pickering emulsions with Bingham-type rheology owing to mechanically reinforced oil–water interfaces. Similarly, Pickering nanoemulsions stabilized by chitosan-coated montmorillonite nanosheets showed enhanced interfacial activity and colloidal stability owing to chitosan-assisted exfoliation \cite{Kang2024AActivity}.

However, platelet-type nanoclays do not universally ensure emulsion stability. For example, Laponite\textsuperscript{\textregistered}-stabilized oil-in-water emulsions exhibited weak stability in the absence of surfactants \cite{Zheng2020EmulsionsApplications}. In contrast, nano-montmorillonite effectively stabilized heavy crude oil–in–water emulsions, with rheology and stability strongly governed by salinity, pH, and temperature \cite{Luo2025RheologicalFluids}. In addition to platelets, tubular nanoclays such as halloysite nanotubes have also been shown to stabilize Pickering emulsions, where lateral interfacial adsorption and bundled shell formation effectively suppressed droplet coalescence \cite{Stehl2020Oil-in-WaterFilterability}.

In contrast to pristine clays, organoclays are organically surface-modified, making them hydrophobic and readily dispersible in oils and organic solvents. For example, layered Na-montmorillonite (Cloisite Na\textsuperscript{+})-stabilized oil-in-water Pickering emulsions exhibited stable, small and narrowly distributed droplets \cite{Yu2021CharacterizationReservoirs}. Liu et al. \cite{Liu2021PickeringConditions} further demonstrated synergistic stabilization using organoclay combined with silica nanoparticles of intermediate hydrophobicity. Individually, the silica nanoparticles failed to stabilize the emulsions because of insufficient interfacial adsorption \cite{Taleb2024ComparativeEmulsion}. In addition to stability, organoclays strongly influence emulsion rheology. Organo-hectorite (Bentone 38)–stabilized water-in-gasoil Pickering emulsions displayed non-Newtonian shear-thinning behavior with a finite yield stress, underscoring the role of clay-induced microstructural networks in controlling flow behavior \cite{Merad2021RheologicalClay}.

Thus, previous studies have indicated that non-spherical particles are effective Pickering emulsion stabilizers. Their higher detachment energies compared to spherical particles, along with their anisotropic interfacial orientation and enhanced surface coverage, particularly for discs and rods, significantly improve emulsion stability \cite{Lisuzzo2022PickeringApplications}. Among non-spherical clays, Garamite-1958\textsuperscript{\textregistered} nano-organoclay shows a strong potential for stabilizing Pickering emulsions. It possesses a unique plate–rod morphology, superior dispersion stability, and inherent thixotropic behavior \cite{Schubel2006CharacterisationReinforcement}. Garamite-1958\textsuperscript{\textregistered} has been explored in various material applications, including polymer and composite systems \cite{Schubel2006CharacterisationReinforcement,Sarathi2007UnderstandingNanocomposites,Rana2021EffectComposites,Ho2006MechanicalNanoclays,Liu2026Garamite-reinforcedDistillation}. However, its potential for forming thixotropic Pickering emulsions for lubrication remains largely unexplored. In addition, studies on thixotropic emulsions, whether nanoclay-based or otherwise, are limited and have mainly been reported in food science \cite{Du2025InvestigationAnalogues,Feng2025Cinnamaldehyde-loadedPerspectives,Lu2021AParticles} and 3D/4D printing \cite{Yu2024VersatileAssemblies,Zhang20254D-PrintableSurfactants}. Even in these cases, tribological investigations are typically restricted to soft contacts, and direct correlations between thixotropy and tribology are rarely established. 

Herein, we introduce Garamite-1958\textsuperscript{\textregistered} nano-organoclay, hereafter referred to as Garamite, as a Pickering emulsion stabilizer for tribological applications. Leveraging its unique plate–rod morphology and inherent thixotropic behavior, we prepared water-in-oil (W/O) Pickering emulsions using sunflower oil, water, and nanoclay, avoiding the use of mineral oils and hazardous additives. Water served as the dispersed phase, and oil served as the continuous phase. Oil was selected as the continuous phase to avoid lubricity deterioration and the limited applications associated with water-continuous systems \cite{Yu2024VersatileAssemblies}. Vegetable oil was selected over mineral oil to ensure environmental compatibility \cite{Wu2017TheEmulsion}. This approach is simple, cost-effective, environmentally benign and scalable. This study systematically investigates rheology, friction, wear and establishes a clear thixotropy–tribology relationship. We further reveal path-dependent and memory-dependent effects in thixotropic lubrication, highlighting the previously untapped functionalities of nanoclay-stabilized emulsions. Our results demonstrate that Garamite-based thixotropic Pickering emulsions deliver effective anti-friction and anti-wear performance under high loads, opening new pathways for smart, sustainable lubricants in the mechanical and materials processing industries.

\section{Results and Discussions}\label{sec2}
\subsection{Emulsion Development and Characterization}\label{subsec2.1}
\subsubsection{Nanoclay characterization}\label{subsubsec2.1.1}

\begin{figure*}[bt]
    \centering     
    \includegraphics[width=\linewidth]{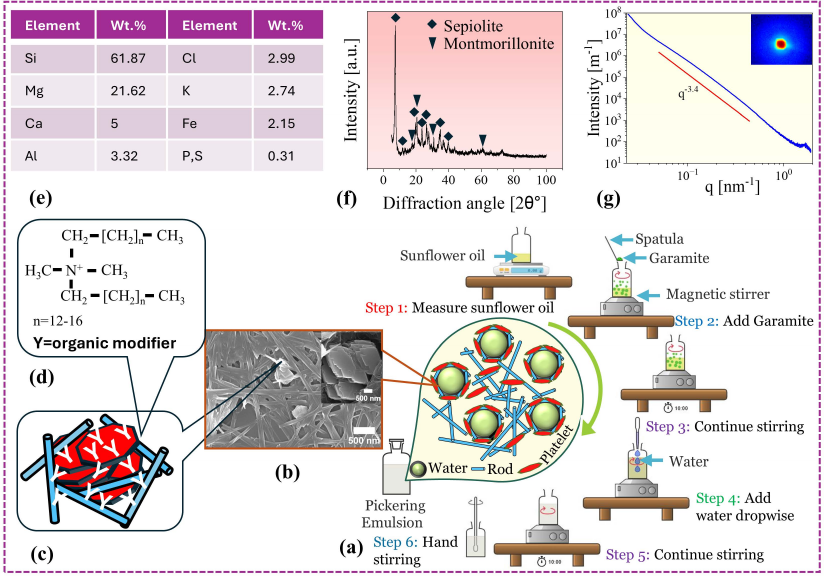}
    \caption{(a) Schematic showing the preparation steps of the thixotropic colloidal emulsion containing Garamite nano-clay, water and sunflower oil. (b) SEM image of Garamite nano-clay revealing its fibrous and platelet-like particle morphology. (c) Illustration of the nano-clay surface structure, showing platelet and rod features modified with alkyl quaternary ammonium organic cations. Redrawn inspired from \cite{BYK2019RheologyAdditives} (d) Chemical structure of the alkyl quaternary ammonium organic modifier. (e) Elemental composition of the nano-clay obtained from XRF analysis. (f) XRD pattern of the nano-clay. (g) I(q) curves obtained by radial integration of the SAXS 2D scattering patterns in the inset.}
    \label{fig:1}
\end{figure*}

\textbf{Figure \ref{fig:1}a} schematically illustrates the emulsion preparation process (details in section \ref{subsec4.2}). Water droplets in the emulsion were co-stabilized by the platelet and rod-like morphology of the nanoclay (chemical structure of nanoclay shown in Figure S1: supporting information (SI)), consistent with similar systems \cite{Lu2021RecentApplications}. The residual Garamite that does not adsorb at the water–oil interface remains in the bulk phase, contributing to a secondary rheological network. The scanning electron microscopy (SEM) micrograph in Figure \ref{fig:1}b reveals the distribution of the clay components. It primarily consists of high-aspect-ratio 1D rods with lengths between $1-5\ \mu\text{m}$ and an average diameter of $\approx$ $32\ \text{nm}$ (refer to Figure S2: SI). The inset in Figure \ref{fig:1}b shows stacked 2D platelets with width of $\approx 800\ \text{nm}$. This structural heterogeneity prevents dense packing and enhances the dispersibility. Figure \ref{fig:1}c shows a schematic of the organoclay arrangement. Both the rods and platelets are modified with alkyl quaternary ammonium cations, the chemical structure of which is shown in Figure \ref{fig:1}d. The elemental composition was determined using X-ray fluorescence (XRF) analysis (Figure \ref{fig:1}e). The high concentrations of Si and Mg indicate a magnesium silicate backbone characteristic of the sepiolite. The minimal Al content distinguishes this material from traditional aluminosilicate bentonites.

%\FloatBarrier

Furthermore, the X-ray diffraction (XRD) pattern (Figure \ref{fig:1}f)  confirms the two-phase (sepiolite and montmorillonite) nature of Garamite, with a prominent basal reflection at $2\theta \approx 7.37^{\circ}$ ($d_{001}\approx1.20\ \text{nm}$). This spacing is characteristic of ammonium-modified organophilic silicates and confirms the successful intercalation. The dominance of the $1.2\ \text{nm}$ phase, alongside diffraction peaks between $20^{\circ}$ and $40^{\circ}$, indicates a structure primarily composed of sepiolite rather than expandable smectite. While XRD identifies these short-range crystalline features, the broad reflections suggest a limited stacking order and partial layer disorder. Small-angle X-ray scattering (SAXS) (Figure \ref{fig:1}g) provides complementary long-range structural data that extend beyond the resolution of XRD. The isotropic circular scattering pattern (inset) indicates that the clay tactoids are randomly oriented \cite{Panda2025AnisotropicActuation}. At high $q$ values, the scattering intensity follows a power-law decay of $q^{-3.4}$, yielding a surface fractal dimension of $D_s = 2.6$ \cite{Teixema1988Small-AngleSystems}. This indicates a complex and rough interface at the nanoscale. A characteristic repeat distance of $ \approx20.2\ \text{nm}$ was obtained from model fitting, which is significantly larger than the interlayer spacing detected by XRD. This length scale is attributed to the interparticle spacing and organic-modifier-induced structuring associated with the clay assembly. Moderate polydispersity (0.199) and low local crystallinity (0.0658) suggest a partially disordered organization. Collectively, these results indicate a hierarchical nanostructure dominated by rod-like structure with platelets dispersed within an organically modified environment.

\subsubsection{Emulsion stability}\label{subsubsec2.1.2}

\begin{figure*}[tb]
    \centering     
    \includegraphics[width=\linewidth]{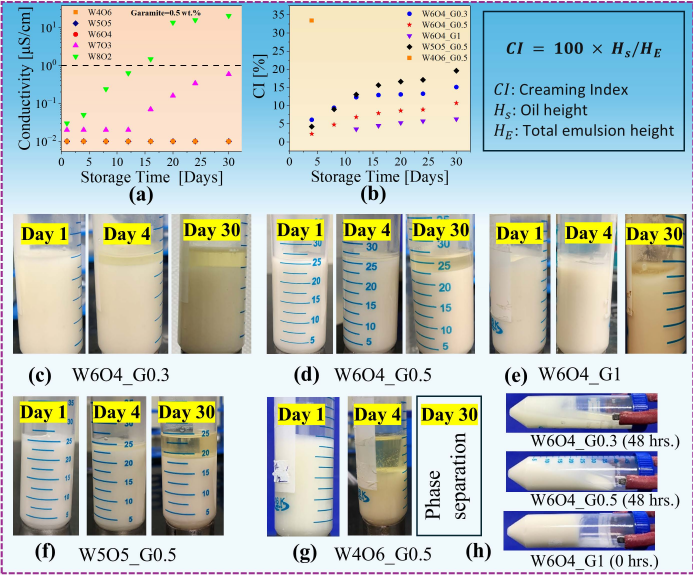}
    \caption{(a) Storage stability of the prepared emulsions evaluated using electrical conductivity measurements, with the Garamite concentration fixed at 0.5 wt.\% for all samples. (b) Storage stability assessed by the creaming index; W7O3 and W8O2 emulsions were excluded because their electrical conductivity values approached 1 $\mu$S/cm, indicating phase instability. (c–g) Representative images of emulsions at different storage times: (c) W6O4\_G0.3, (d) W6O4\_G0.5, (e) W6O4\_G1, (f) W5O5\_G0.5, and (g) W4O6\_G0.5. Based on the stability analysis, the W5O5 and W4O6 systems were further excluded, and subsequent experiments were conducted using the W6O4 emulsion with Garamite concentrations of 0.3, 0.5, and 1 wt.\%. (h) Time-dependent images of W6O4 emulsions within 48 h after tilting the storage bottle to $90^\circ$ to observe gelation.}
    \label{fig:2}
\end{figure*}

Following nanoclay characterization, a set of five emulsions (W8O2, W7O3, W6O4, W5O5, and W4O6) with the corresponding water–oil weight ratios (4:1, 7:3, 3:2, 1:1, and 2:3), as detailed in section \ref{subsec4.2}, was systematically designed to evaluate formulation-dependent stability for long-term storage. The Garamite content in these emulsions was 0.5 wt.\%. Initially, the emulsions were screened using electrical conductivity measurements over 30 days (\textbf{Figure \ref{fig:2}a}). The W4O6, W5O5, and W6O4 systems exhibited conductivities on the order of (${\sim}10^{-2}\ \mu\text{S/cm}$) throughout the testing period. This confirms that the oil remained in the continuous phase, effectively insulating dispersed water droplets. In contrast, the W7O3 system showed a marked increase in conductivity after 12 d, reaching approximately $1\ \mu\text{S/cm}$ by day 30. This increase signifies phase separation, as the coalesced water phase establishes a conductive path to the probe. The water-rich W8O2 system demonstrated this instability from day one, with conductivity steeply exceeding $1\ \mu\text{S/cm}$ within 15 days. Consequently, W7O3 and W8O2 were excluded from further studies because of their insufficient kinetic stabilities.

The storage stability of the remaining emulsions was then assessed using the creaming index (CI) (Figure \ref{fig:2}b). For the W6O4 system, the Garamite content was varied to 0.3, 0.5, and 1 wt.\%, denoted as G0.3, G0.5, and G1, respectively. All the W6O4 compositions showed a gradual increase in the CI. G1 exhibited the lowest CI. This suggests relatively high colloidal stability by reinforcing the interfacial Pickering shield and the bulk rheological network. The W5O5 and W4O6 systems (0.5 wt.\% Garamite) were also evaluated. W5O5 showed a more rapid increase in CI than W6O4. The W4O6 system failed within four days, indicating phase separation, likely because of an insufficient water-to-oil ratio to maintain a stable droplet morphology. The corresponding visual snapshots at different storage times are shown in Figure \ref{fig:2}(c–g) and Figure S2 (SI). In conclusion, the W6O4 emulsion exhibited superior stability. Stability arises from the formation of a rigid barrier by solid particles at the oil–water interface. Irreversible nanoparticle adsorption, driven by high desorption energy, forms strong steric barriers that suppress droplet coalescence, deformation, and Ostwald ripening \cite{Lisuzzo2022PickeringApplications}. In addition, the rod-like morphology of sepiolite promotes network formation, further enhancing the dispersion and emulsion stability \cite{Liu2026Garamite-reinforcedDistillation}. The W6O4 emulsion was selected for all subsequent analyses. All experiments and analyses were conducted within two days of emulsion preparation. Physical aging was visually monitored using the vial-tilting method. The emulsion was placed in a transparent bottle and tilted at $90^\circ$ at regular intervals over 48 h. The G0.3 system maintained its fluidity throughout the observation period, whereas G0.5 exhibited only partial flow at 48 h. In contrast, G1 displayed an immediate loss of mobility after preparation (Figure \ref{fig:2}h); additional images are shown in Figure S3:SI). These observations indicate that increasing the Garamite concentration promotes the formation of a progressive structure. Over time, the residual clay forms a particle network in the continuous phase. The droplets become trapped within this network, and the system evolves from a flowing emulsion to a Pickering gel-type emulsion as the Garamite concentration increases.

%\FloatBarrier

\subsubsection{Emulsion characterization}\label{subsubsec2.1.3}

\begin{figure*}[tb]
    \centering     
    \includegraphics[width=\linewidth]{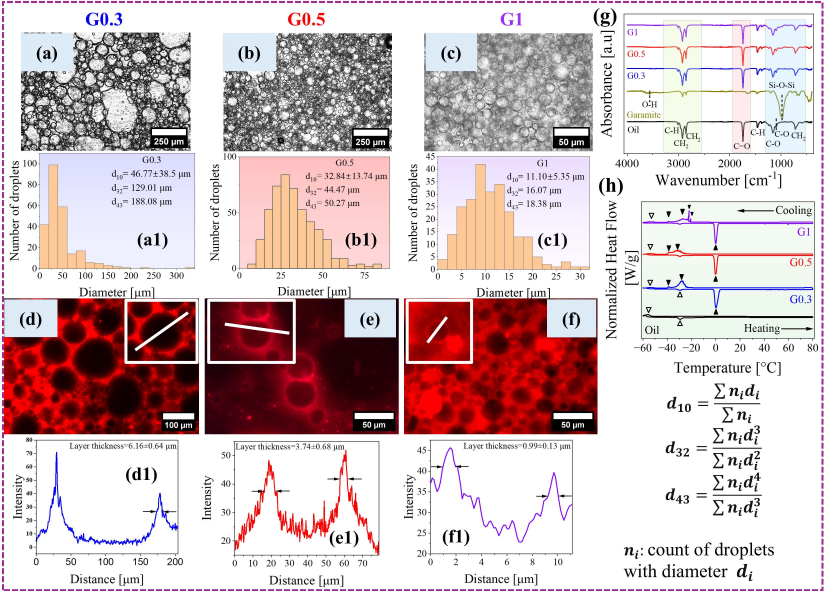}
    \caption{Optical micrographs and corresponding droplet size distribution histograms of emulsions (a, a1) G0.3, (b, b1) G0.5, and (c, c1) G1. Fluorescence microscopy images with associated intensity profiles measured along the lines indicated in the insets (zoomed droplet views) for estimating the nanoclay interfacial layer thickness: (d, d1) G0.3, (e, e1) G0.5, and (f, f1) G1. (g) FTIR spectra of sunflower oil, Garamite nano-clay, and the as-prepared emulsions (h) DSC heating and cooling thermograms of sunflower oil and the as-prepared emulsions. Micrographs (a) and (b) use the same scale, while (c) is shown at higher magnification to resolve the significantly smaller and more numerous droplets at G1 concentration. $d_{10}, d_{43}$ represents number and volume weighted mean diameters whereas $d_{32}$ represents Sauter mean diameter.}
    \label{fig:3}
\end{figure*}

The optical micrographs and droplet size distributions in \textbf{Figure \ref{fig:3}} illustrate the structural evolution of W6O4 emulsions. A clear transition from a coarse, polydisperse system in G0.3 to a fine and uniform dispersion in G1 is evident. In the G0.3 system (Figure \ref{fig:3}a), large non-spherical droplets dominate the morphology, with smaller spherical droplets adhering to their surfaces. This irregular shape suggests that low clay concentrations provide insufficient interfacial armoring, leading to partial coalescence and structural relaxation. In contrast, the G0.5 system (Figure \ref{fig:3}b) exhibits a relatively uniform distribution. The droplets are primarily spherical, indicating that 0.5 wt.\% Garamite provides enough interfacial coverage to maintain structural integrity. By G1 (Figure \ref{fig:3}c), the increased availability of nanoclay at the water–oil interface resulted in a significantly smaller and more crowded droplet morphology.

The statistical parameters in Figures \ref{fig:3}a1–c1 quantify this evolution. The number-weighted mean diameter ($d_{10}$) decreases sharply from $46.77 \pm 38.5\ \mu\text{m}$ in G0.3 to $32.84 \pm 13.77\ \mu\text{m}$ and $11.10 \pm 5.35\ \mu\text{m}$ for G0.5 and G1, respectively. The high standard deviation in G0.3 confirms its broad polydispersity. The surface-weighted/ Sauter mean diameter ($d_{32}$) also decreased from $129.01\ \mu\text{m}$ in G0.3 to $16.07\ \mu\text{m}$ in G1. This parameter is particularly significant as it directly relates to the total interfacial area requiring clay stabilization \cite{Stehl2020Oil-in-WaterFilterability}. The volume-weighted mean diameter ($d_{43}$), which is highly sensitive to large droplets, decreased from $188.08\ \mu\text{m}$ (G0.3) to $18.38\ \mu\text{m}$ (G1). This trend indicates a significant decrease in the volume fraction of coalesced water. Higher Garamite loading enhances steric hindrance, preventing droplet fusion and reducing polydispersity. Ultimately, the narrow distribution and lower $d_{43}$ values of the G1 system indicate a superior emulsion quality and enhanced kinetic stability against creaming.

Figures \ref{fig:3}d–f and  \ref{fig:3}d1–f1 show the fluorescence microscopy images and corresponding intensity profiles of the as-prepared emulsions. The Nile Red hydrophobic dye yielded fluorescent signals from the continuous oil phase containing Garamite, confirming the successful preparation of W/O Pickering emulsions. The bright fluorescent halos at the water–oil interface represent protective Pickering shields. The shell thickness was quantified using the full width at half-maximum of the intensity peaks across the droplet boundaries \cite{Kong2025ElucidatingMayonnaise}. A clear thinning of the interfacial shell occurs as Garamite concentration increases from 0.3 to 1 wt.\%. The G0.3 system exhibits the thickest shell at 6.16 $\pm$ 0.64 $\mu\text{m}$, suggesting the formation of loosely packed clay aggregates. This correlates with the large, non-spherical droplets observed in optical microscopy, where insufficient particle dispersion fails to create a cohesive barrier. As loading increases to 0.5 wt.\%, the shell thickness reduces to 3.74 $\pm$ 0.68 $\mu\text{m}$, indicating more compact nanoclay packing. By G1, the shell reaches its thinnest and most uniform state at 0.99 $\pm$ 0.13 $\mu\text{m}$. This sub-micron thickness provides superior steric hindrance by effectively jamming the interface. 

The Fourier transform infrared (FTIR) spectra of the oil, Garamite, and emulsions are shown in Figure \ref{fig:3} (g). Pure Garamite exhibits a characteristic broad band around 3575 cm\textsuperscript{-1} associated with O-H stretching and a dominant sharp peak at 990 cm\textsuperscript{-1} corresponding to the Si-O-Si stretching vibrations. The oil phase displays prominent peaks in the 2850–3000 cm\textsuperscript{-1} region, assigned to C-H stretching ($\text{CH}_2$ and $\text{CH}_3$ groups), and a strong C=O stretching band at 1743 cm\textsuperscript{-1} related to the ester groups in the oil matrix. As observed in the emulsion spectra, the absorption bands primarily resembled those of the pure oil phase. This dominance is attributed to the significantly higher volume fraction of oil compared to the nanoclay content. The lack of peak shifting in the C=O or O-H regions in the case of emulsions indicates the absence of strong chemical grafting or covalent bonding between Garamite and oil phase. These results suggest that the formation of the W/O Pickering emulsion is governed by physical forces rather than chemical interactions.
 
%\FloatBarrier

The phase-transition behavior of the oil and emulsions, using differential scanning calorimetry (DSC), as a function of Garamite loading is summarized in \textbf{Table \ref{table:2}} and Figure \ref{fig:3} (h). DSC heating and the cooling thermograms of Garamite nano-clay are shown in Figure S4: SI. While the fusion points ($T_f$) of bulk water ($0 ^\circ$C) and oil ($-29.3  ^\circ$C) remain constant, their enthalpies exhibit a distinct non-monotonic trend. Notably, the water fusion enthalpy ($\Delta H_f$) reached a minimum at 0.5 wt.\% loading ($153.76\text{ J g}^{-1}$), compared to $200.68\text{ J g}^{-1}$ for G0.3 and $158.16\text{ J g}^{-1}$ for G1. This enthalpy depression at 0.5 wt.\% signifies a high proportion of non-freezable or interfacially immobilized water. Furthermore, the oil fusion enthalpy peaked at 0.5 wt.\% ($11.2\text{ J g}^{-1}$), suggesting that this intermediate loading most effectively bridges both phases, avoiding the localized aggregation observed at 1 wt.\%.

\begin{table}[h!]
\centering
\renewcommand{\arraystretch}{1.3}
\setlength{\tabcolsep}{4pt}
\caption{Fusion ($T_f$) and crystallization ($T_c$) temperatures with corresponding enthalpies ($\Delta H_f$, $\Delta H_c$) of oil and water in emulsions.}
\resizebox{\linewidth}{!}{
\label{table:2}
\begin{tabular}{c c c c c c c c c}
\hline
\multirow{2}{*}{Sample} 
& \multicolumn{2}{c}{$T_f$ ($^{\circ}$C)} 
& \multicolumn{2}{c}{$T_c$ ($^{\circ}$C)} 
& \multicolumn{2}{c}{$\Delta H_f$ (J g$^{-1}$)} 
& \multicolumn{2}{c}{$\Delta H_c$ (J g$^{-1}$)} \\

\cline{2-9}
& Oil & Water & Oil & Water & Oil & Water & Oil & Water \\
\hline

Oil 
& -29.3 & -- 
& -56.1 & -- 
& 22.8 & -- 
& 25.79 & -- \\

G0.3 
& -29.3 & 0 
& -54.17 & [-38.67, -27.15] 
& 10.4 & 200.68 
& 13.36 & 162.4 \\

G0.5 
& -29.3 & 0 
& -53.91 & [-38.45, -31.78] 
& 11.2 & 153.76 
& 15.39 & 124.32 \\

G1 
& -29.3 & 0 
& -52.93 & [-38.78, -27.47, -22.13, -20.62] 
& 10.24 & 158.16 
& 13.01 & 125.6 \\

\hline
\end{tabular}
}
\end{table}

Cooling thermograms validated the optimal stability. Oil crystallization temperatures ($T_c$) shifted toward higher values with increasing loading, from $-54.17 ^\circ$C (G0.3) to $-52.93 ^\circ$C (G1), confirming Garamite’s role as a heterogeneous nucleating agent. For the water phase, G1 exhibited four distinct peaks, reflecting high droplet compartmentalization and heterogeneous nucleation at particle-rich interfaces \cite{Thompson2001ConfinementMelting, Morishige1999FreezingBehavior}. However, the second crystallization peak for G0.5 occurred at the lowest temperature ($-31.78^\circ$C) compared with G0.3 ($-27.15^\circ$C) and G1 ($-27.47^\circ$C), indicating that water in the G0.5 system requires the highest degree of supercooling, confirming superior interfacial protection and smaller freezable domains. Interestingly, although microscopy identified G1 as having the smallest mean droplet size, the DSC data identified G0.5 as the system with the highest interfacial confinement and lowest water crystallization enthalpy ($\Delta H_c$: $124.32\text{ J g}^{-1}$). The slight enthalpy increase in G1 suggests that excessive clay loading leads to particle-particle aggregation in the continuous phase, reducing the efficiency of the water-particle interaction. Consequently, whereas G1 displays a finer visual morphology, G0.5 represents the thermodynamically optimal balance.

The desorption energy $\Delta E_{des}$ of rod-shaped nanoparticles from the oil-water interface is given as follows \cite{Lisuzzo2022PickeringApplications}:
\begin{equation}
        \Delta E_{\mathrm{des}} =  \gamma_{o/w} lq(1-|\cos\theta| ))
\end{equation} 
where $\gamma_{\mathrm{o/w}}$ is the interfacial tension at the oil-water interface, $\theta$ is the three-phase (oil, water, and Pickering solid) contact angle ($> 90^\circ$ for W/O emulsions), and $l$ and $q$ are the lateral and longitudinal semi-axial lengths of the nanorod clay particle, respectively. The $\gamma_{\mathrm{o/w}}$ values for G1, G0.5, and G0.3 were 55.44, 83.84, and 34.81 mN/m, respectively. Because $\Delta E_{\mathrm{des}}$ is proportional to $\gamma_{\mathrm{o/w}}$, assuming other variables are constant, a higher $\Delta E_{\mathrm{des}}$ for G0.5 implies superior emulsion stability, which is consistent with the DSC findings.

\subsection{Rheological Investigation}\label{subsec2.2}

\begin{figure*}[tb]
    \centering     
    \includegraphics[width=\linewidth]{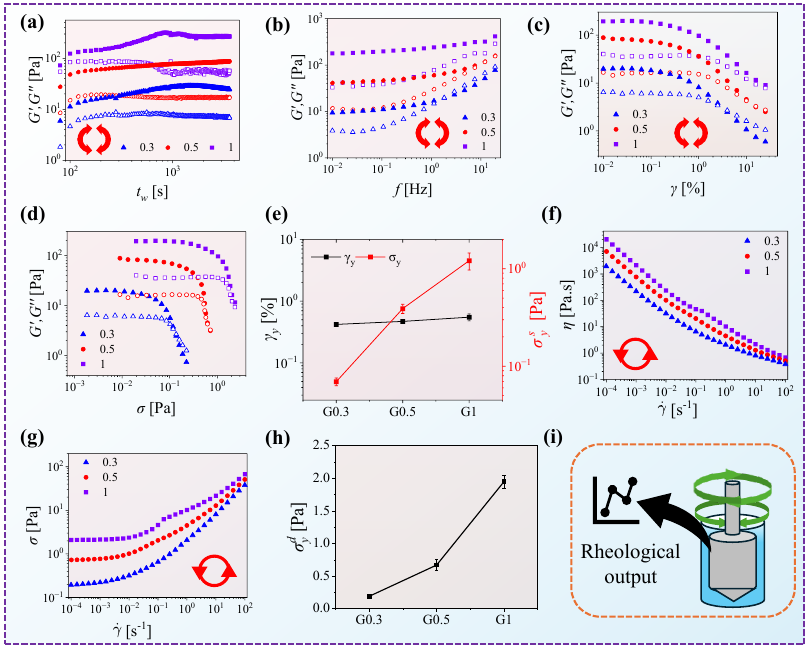}
    \caption{ For the W6O4 system at Garamite concentrations of 0.3, 0.5, and 1 wt.\%.(a) The evolution of elastic modulus $G'$ (solid symbol) and viscous modulus $G''$ (open symbol) with waiting time ($t_w$) at  0.1 \% strain amplitude ($\gamma_0 $) and 0.1 Hz frequency   (b) $G'$ and $G''$  evolution as a function of frequency [Hz] for 0.1 \% strain amplitude.(c) and (d) are elastic modulus $G'$  and viscous modulus $G''$  plotted as the function of applied dynamic strain amplitude sweep and it's resultant stress output. (e) Calculated yield strain and static yield stress from the dynamic amplitude sweep experiment. (f) and (g) are the plots of  viscosity and shear stress as a function of applied shear rate $\dot{\gamma}$ in the flow sweep experiment. In (f), for all the systems, shear thinning behavior is observed. (h) Calculated dynamic yield stress from the flow sweep experiment. (i) Schematic of cup and cylindrical geometry used. The insets in Fig (a),(b),(c) show that the test is oscillatory, while the inset in \ref{fig:4}a,b shows the test is rotary.}
    \label{fig:4}
\end{figure*}

We perform a detailed rheological investigation, and the associated behavior is described in \textbf{Figure \ref{fig:4}}. We first perform small amplitude oscillatory shear experiments, subsequent to the stoppage of shear melting, to assess the evolution of dynamic moduli as a function of time. The corresponding results are depicted in Figure \ref{fig:4}a), which indicates that across all concentrations, the storage modulus $G'$ exceeds the loss modulus $G''$, demonstrating predominantly solid-like behavior. The increase in $G'$  with waiting time suggests physical aging, which is a natural tendency for systems that are out of their equilibrium state to reach a more thermodynamically favorable microstructural rearrangement (progressively lower free energy states) during the course of time \cite{Joshi2014DynamicsGels, Joshi2015AMaterials}. This physical aging is more pronounced for G1 ($t^{0.6}$) followed by G0.3 ($t^{0.3}$) and G0.5 ($t^{0.1}$), indicating  that G0.5 system exhibits relatively stable modulus growth without excessive stiffening. At the end of the waiting time period ($t_w$), the plateau modulus ($G'$) achieved was 270, 87, and 25 Pa for G1, G0.5, and G0.3, respectively, which again shows that the G1 system becomes stiffer, possibly due to the formation of an interconnected bridged network between the excess Garamite particles around the   ter droplets, as shown in Figure \ref{fig:1}a, thereby hindering their movement and strengthening the system. 

Figure \ref{fig:4}b illustrates the frequency sweep results, where an increase in both dynamic moduli ($G'$ and $G''$) is observed with increasing Garamite wt.\%. Eventually, $G'$  becomes independent of frequency for G1. Such weak frequency dependence is characteristic of solid-like behavior and elastic network formation, which is caused by droplet particle bridging network at high concentrations of Garamite particles \cite{Munro2022YieldingFractions}. For both G0.5 and G0.3, minima were observed in $G''$ at 0.0258 and 0.064 Hz, respectively, which is indicative of soft glassy dynamics \cite{Kaushal2014LinearMaterials, Keane2025UniversalMaterials}. A crossover point is observed only for the G0.3 system in the explored frequency regime at $\omega$ = 1.77 rad s\textsuperscript{-1}, which signifies a frequency-driven transition from the elastic to the glassy viscous regime.

Figure \ref{fig:4}c illustrates the evolution of the dynamic moduli ($G'$,$G''$) as a function of the applied strain amplitude ($\gamma_0$). Both $G'$ and $G''$ increase as the Garamite loading increases. The linear viscoelastic regime (where $G'$ and $G''$ remain independent of the applied strain amplitude ($\gamma_0$)) extends approximately up to  0.2 \% strain for all Garamite wt.\%. $G'$ and $G''$ as a function of the induced stress in the system is shown in Figure \ref{fig:4}d, which shows that increasing the Garamite wt.\% increases the stress induced in the system, due to a more interconnected network of residual Garamite particles in the oil phase matrix. The static yield stress ($\sigma_y^s$) is defined as the minimum stress required to initiate flow in a material under rest conditions \cite{Joshi2018YieldAgeing}. Its value, along with the yield strain ($\gamma_y$), is obtained by taking the intersection point of the low- and high-amplitude asymptotes in their respective moduli function plots (Figure \ref{fig:4}c and d). Their values are illustrated in Figure \ref{fig:4}e, which shows a noticeable increase with increasing Garamite wt.\%. Owing to the almost  same linear regime for all systems, the $\gamma_y$ shows only a slight increase. $\sigma_y^s$ on the other hand increases significantly, which could be due to the material stiffening provided by the residual Garamite particles interconnected by the bridged network in the continuous oil phase covering the water droplets. This type of behavior is observed in soft colloids \cite{Ciccone2022AFormulations, Majumder2026RheologyMethod, Agrawal2024InterplayRheology}. 

%Additionally, for such soft colloidal systems, Nordstrom et al. showed that $\sigma_y$ scales quadratically $(\phi-\phi_c)^2$  and $\gamma_y$ scales linearly $(\phi-\phi_c)$ with the distance from the jamming point governed by the dispersed phase volume fraction ($\phi$) \cite{Nordstrom2010MicrofluidicJamming} where jamming refers to the transition from a flowing state to a mechanically rigid or arrested state that occurs as a result of crowding, packing, or interactions among particles, droplets, or other structural elements in the material matrix.  

In Figure \ref{fig:4}c, the shifting of the ($G'$ – $G''$) crossover to a higher strain value with increasing Garamite wt.\% suggests that a larger deformation is required to disrupt the strengthened internal structure and initiate irreversible flow in the system.

%\FloatBarrier

Figure \ref{fig:4}f shows the viscosity as a function of shear rate, where shear-thinning behavior was observed, which became more pronounced with increasing Garamite wt.\%, with viscosity decreasing over several orders of magnitude in the explored shear rate range. Higher Garamite loading increases the low shear viscosity owing to a stronger droplet-particle network bridging, whereas at high shear, this structure is progressively disrupted, and close viscosity values are observed for all systems. Shear thinning enables the lubricant to slip between the tribo-pair and provides lubrication through film formation.

Figure \ref{fig:4}g shows the shear stress as a function of the shear rate. A monotonic decrease in stress was observed with decreasing shear rate, and at a small shear rate, a plateau was observed, indicating the presence of a dynamic yield stress ($\sigma_y^d$), which is the minimum stress required to maintain flow once it has been commenced. Overall, the stress induced in the material was higher for higher Garamite concentrations.
 Figure \ref{fig:4}h shows that the dynamic yield stress values, calculated from the stress curve plateau value at low shear rate \cite{Bhattacharyya2023OnMaterials}, increase with the increase in  Garamite concentration, suggesting a more bridged network formation of residual Garamite particles in the continuous oil phase matrix left after adsorption into water droplets, strengthening the system, as shown in Figure \ref{fig:4}a.
 It is interesting to note that generally $\sigma_y^s$ $>$ $\sigma_y^d$ and the same has also been observed for 0.5 and 0.3 wt.\%, but for 1 wt.\%, the value of $\sigma_y^d$ is slightly higher, and the possible reason for this is due to a greater structure build-up rate (aging) than structure breaking rate (rejuvenation) at lower values of $\dot{\gamma}$ which is commonly observed for thixotropic soft glassy materials \cite{Joshi2014DynamicsGels, Joshi2025LinearMaterials}.

\begin{figure*}[tb]
    \centering     
    \includegraphics[width=\linewidth]{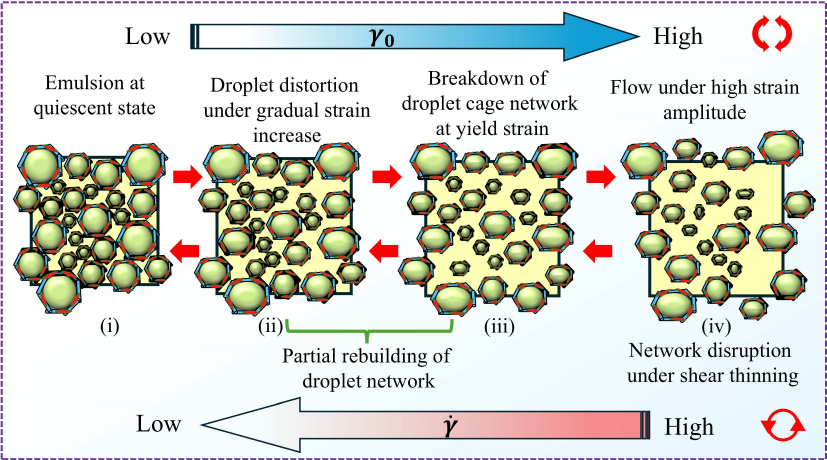}
    \caption{ Microstructure evolution during dynamic oscillatory amplitude sweep experiment at a fixed frequency with  $\gamma_0$ increasing from left to right, shown on top, and steady state flow sweep experiment with $\dot{\gamma_0}$ decreasing from right to left shown in bottom. The symbols at the top and bottom right corner shows the oscillatory and rotary nature of the experiment.}
    \label{fig:5}
\end{figure*}

Furthermore, the evolution of the microstructure during the dynamic oscillatory strain sweep and steady-state flow sweep experiments is illustrated in  \textbf{Figure \ref{fig:5}}, with an attempt to understand the microscopic origin of  $\sigma_y^s$ and  $\sigma_y^d$.

%\FloatBarrier
 
The emulsion droplets, as shown in Figure \ref{fig:5}(i), are surrounded by their neighboring droplets, forming a cage-like network during the quiescent state, which resists flow up to the linear viscoelastic (LVE) regime. As the applied strain amplitude increases beyond the LVE regime, the droplet network distorts to some extent to free up the matrix volume (Figure \ref{fig:5}(ii)). A further increase in the strain amplitude weakens this caged network and ultimately causes it to break (Figure \ref{fig:5}(iii)) at the yield point, $\gamma_y$ . At a strain beyond this yield point, water droplets with an elastic nature elongate into an ellipsoidal shape (Figure \ref{fig:5}(iv)), which can slide past each other, and liquid-like flow starts after the crossover point of $G'$ and $G''$ as depicted in Figure \ref{fig:4}c. 
 
In the case of the steady-state flow sweep experiment (high to low $\dot{\gamma_0}$ ), a similar type of microstructure (Figure \ref{fig:5}(iv)) exists at a high shear rate. As the shear rate gradually decreases, the droplets, owing to their elastic nature, restore their original shape up to a certain limit, and partial rebuilding of the droplet network occurs owing to the thixotropic nature of the emulsion imparted by Garamite (Figure \ref{fig:5}(iii \& ii)). Under steady shear, particularly at very low shear rates, the system maintains a continuously restructuring, transiently connected network, as a fully percolated network cannot be sustained. This near balance between the formation of aggregates and shear-induced disruption of the same leads to a time-averaged connectivity that manifests as a stress plateau, interpreted as a dynamic yield stress.
 
 Overall, the rheological study revealed that the W6O4 emulsion stabilized by different Garamite concentrations exhibited thixotropic behavior, where the viscosity decreased upon application of a high shear rate, as shown in \ref{fig:4}f. A rebuilding of the microstructure network occurred with the evolution of the dynamic moduli ($G'$,$G''$) with waiting time ($t_w$), which became more pronounced with increasing Garamite loading percentage, as shown in Figure \ref{fig:4}a.
 In addition to this Garamite loading has a significant effect on the yield stress of the system which shows an increase in both (static and dynamic) yield stress value upon increasing it's loading percentage.
 
\subsection{Frictional Performance}\label{subsec2.3}

\begin{figure*}[tb]
    \centering     
    \includegraphics[width=\linewidth]{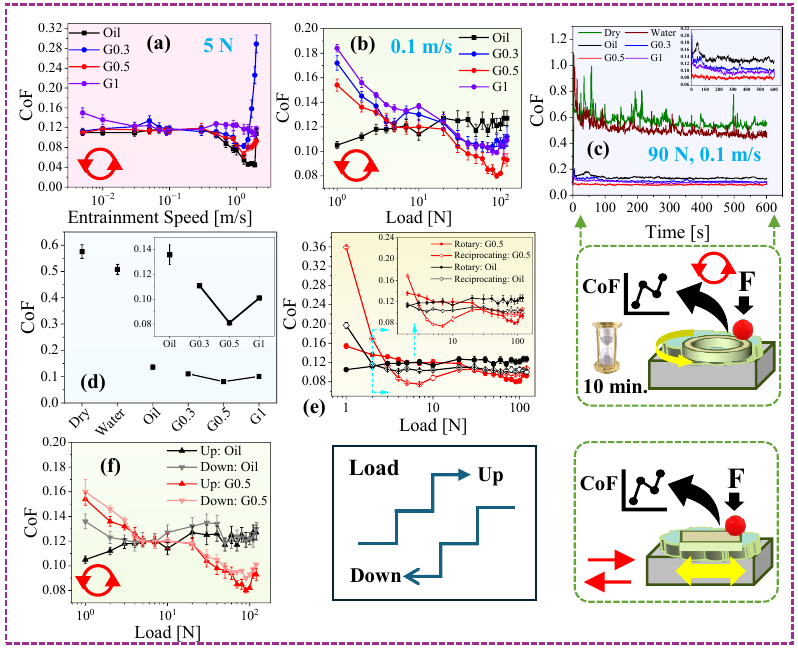}
    \caption{(a) CoF as a function of entrainment speed under a constant normal load of 5 N (b) CoF as a function of applied normal load at a fixed sliding speed of 0.1 m s\textsuperscript{-1} (c) Evolution of CoF with test duration under submerged lubrication at a load of 90 N and a sliding speed of 0.1 m s\textsuperscript{-1} in rotary (unidirectional) sliding. (d) Average CoF values derived from the steady-state regime. (e) CoF versus normal load for the G0.5 emulsion at 0.1 m s\textsuperscript{-1} under reciprocating and rotary sliding modes. (f) CoF versus load for the G0.5 emulsion under stepwise loading sequences, with load increased from low to high and decreased from high to low. Tests were conducted on self mated AISI304 tribo-pair. Error bars represent the standard deviation; for points where error bars are not visible, the standard deviation is within the symbol size.}
    \label{fig:6}
\end{figure*}

To ensure consistency across lubrication conditions, the specimens were polished to a uniform root mean square roughness ($R_q$) of $18.8 \pm 1.58 \text{ nm}$. The tribological performance was evaluated using unidirectional rotary and linear reciprocating tests in a ball-on-disc configuration. Given its wide industrial application, self-mated AISI 304/AISI 304 stainless-steel specimens were used as tribopairs. \textbf{Figure \ref{fig:6}a} illustrates the mean coefficient of friction (CoF) as a function of entrainment speed (0.005-2 m s\textsuperscript{-1}) under a 5 N load in rotary sliding mode. At low speeds ($< 0.01$ m s\textsuperscript{-1}), the G1 system exhibits a relatively higher CoF, correlating with its high dynamic yield stress and immediate gelation, which resists initial flow. Between 0.01 and 0.3 m s\textsuperscript{-1}, the frictional profiles of all lubricants overlap. Beyond 0.3 m s\textsuperscript{-1}, the CoF for the base oil, G0.3, and G0.5 reaches a minimum before rising. Notably, G0.3 shows a steep increase in the CoF at higher speeds. This suggests that the loosely armored droplets in G0.3, characterized by a thick (6.16 \textmu m) but low-density shell, undergo shear-induced coalescence. At high shear rates, these droplets likely rupture and enter the tribo-contact, causing oil film failure. Conversely, the G1 system remains stable throughout the speed range owing to its jammed sub-micron Pickering shield, which provided superior steric hindrance. However, at extreme velocities ($> 1$ m s\textsuperscript{-1}), the base oil outperforms the emulsions, indicating that clay-stabilized droplets may introduce additional resistance. However, G0.5 exhibited the optimal performance. Further increasing the Garamite concentration to 3 wt.\% (Figure S5:SI) shows that friction decreases up to G0.5 and increases thereafter, confirming G0.5 as optimal formulation.

%\FloatBarrier

The mean CoF variation as a function of the applied load (1-120 N, corresponding to the initial mean Hertzian contact pressures of 0.43–2.1 GPa) is shown in Figure \ref{fig:6}b. For the base oil, the CoF increases with load and stabilizes at $\sim$ 0.12–0.13 beyond 20 N. In contrast, all emulsions exhibit higher CoF than the oil at low loads ($\le$ 5 N). Among the emulsions, G0.5 exhibited the lowest friction, while G1 exhibited the highest. As the load increases ($>$ 5 N), all emulsions demonstrate a consistent decrease in the friction trend up to 90 N, followed by a gradual increase. A significant CoF reduction was achieved by G0.5 compared to the base oil at 90 N. Even at a high initial Hertzian contact pressure of $\approx$ 1.9 GPa, the G0.5 system resisted structural breakdown. This suggests that its thermodynamically optimal structure effectively prevents droplet rupture and water release in the contact zone. Although G0.3 and G1 performed similarly at high loads, G0.5 remained the superior lubricant. These results highlight the high load-bearing capacity and structural robustness of Garamite-stabilized Pickering emulsions, which are essential for applications such as metalworking and heavy machinery.

To understand these friction variations, we initially analyzed the results by looking at the classical Stribeck lubrication regimes \cite{Qin2025TribologyMaterials}. Using the Hamrock-Dowson formula \cite{Hamrock1977IsothermalResult} to determine the minimum film thickness, we calculated the lambda ratio ($\lambda$). The detailed analytical approach is described in \cite{Kumar2025Nano-silicaLubricant}. In Figure \ref{fig:6}a, the system remained within the boundary lubrication regime up to 1.3 m s\textsuperscript{-1}, transitioning into the mixed regime and continuing in this regime until a maximum speed of  2 m s\textsuperscript{-1}. Conversely, the data in Figure \ref{fig:6}b remains entirely within the boundary regime. Notably, because the Hersey number is inversely proportional to the load, the horizontal axis should be interpreted as moving toward the boundary regime from left to right. Traditional Stribeck analysis provides only a partial understanding of these transitions, as the available models for calculating film thickness assume Newtonian fluids. Our emulsions are thixotropic and stabilized by non-spherical particles, and their high water content places them between isoviscous and piezoviscous fluids. Moreover, conventional models consider rolling–sliding contacts, whereas our tests involved pure sliding. Thus, existing frameworks developed for single-phase lubricants oversimplify the contact mechanics and fail to capture the regime transitions of these emulsions \cite{Wang2025PickeringPerformance}. 

To understand the frictional variations in these multiphase emulsions, interfacial structuring and shear-dependent behaviors must also be considered. As shown in Figure \ref{fig:6}a, the friction initially decreased with increasing sliding velocity, likely due to the enhanced entrainment and inlet structuring of the emulsion. High aspect-ratio residual rods and droplets accumulate near the contact inlet, locally increasing the effective viscosity and promoting the formation of thicker films. However, beyond $\approx$ 1.3 m s\textsuperscript{-1}, intense shear disrupts the thixotropic network and reduces the inlet residence time, limiting particle accumulation and destabilizing the lubricating film, thereby increasing asperity interactions and friction. As shown in Figure \ref{fig:6}b, at a constant sliding speed, the friction decreases owing to the load-induced stabilization of the interfacial layer. The oil forms an adsorbed boundary film, while the rods and platelets align along the shear direction, creating a compact load-bearing structure that maintains surface separation. The thixotropic emulsion further enables reversible structure recovery, ensuring efficient lubricant supply within the contact. Consequently, the optimally structured emulsion (G0.5) formed a more robust load-supporting film than oil alone, resulting in lower friction.

Next, to assess the frictional stability, the CoF was monitored for 10 min at 90 N and 0.1 m s\textsuperscript{-1} for oil and emulsions submerged surface (Figure \ref{fig:6}c). Dry and water lubrication were also tested for comparison. Oil exhibited larger CoF fluctuations than the emulsions, whereas G0.5 showed the smoothest evolution, as highlighted in the inset of Figure \ref{fig:6}c. The mean CoF values (Figure \ref{fig:6}d) show that oil has the highest CoF, which decreases for G0.3 and reaches a minimum for G0.5, followed by a slight increase for G1, indicating an optimal composition of G0.5. Overall, G0.5 achieved $\approx$ 41\% and $\approx$ 84\% lower friction compared with oil and water lubrication, respectively.

The effect of motion direction on friction was examined by performing linear reciprocating tests on the G0.5 emulsion over the same load range used in the rotary tests (Figure \ref{fig:6}e). Tests with oil were also conducted to isolate the contribution of the emulsion structure. Similar load-dependent trends were observed for both motions in the case of oil, with a slightly lower CoF in reciprocating sliding. For G0.5, the rotary tests showed a minimum CoF at 90 N, as discussed earlier. However, in reciprocating motion, a double-minimum behavior appeared, with the first minimum at 7 N and the second at 90 N. Notably, the second minimum exhibited a slightly higher CoF than that of the rotary sliding. The shift of the minimum friction point to a lower load in reciprocating motion highlights the sensitivity of the emulsion lubrication mechanism to the sliding direction. 

The double-minimum friction profile observed in reciprocating sliding likely originates from the shear-responsive structure and confinement behavior of the emulsion. In reciprocating motion, periodic velocity reversal introduces transient low shear intervals that allow partial recovery of the thixotropic clay–droplet network. At low loads ($\approx$ 7 N), this structural rebuilding promotes the formation of localized micro-reservoirs, where rods and platelets bridge surface asperities and retain oil–water droplets within the contact. This early structural padding stabilizes the boundary film and leads to the first friction minimum, which is absent in continuous rotary sliding. With increasing load, stronger confinement progressively compacts the rod–droplet assemblies and flattens the armored Pickering droplets into dense pancake-like interfacial layers, forming a robust load-bearing tribofilm. Simultaneously, clay particles can cap surface asperities and smoothen the contact path, while the confined water phase dissipates flash heat and maintains oil viscosity. These synergistic effects produced a second minimum near $\approx$ 90 N. The earlier appearance of the first minimum in reciprocating motion reflects partial structural recovery under cyclic shear, facilitating contact replenishment and effective lubrication at lower loads than in rotary sliding.

The increase in the CoF beyond 90 N (Figure \ref{fig:6}b) suggests structural disruption within the emulsion. This raises the question of whether the structure can recover once the load is reduced, particularly considering the possible interference from wear debris. To examine this, the load was reversed from high to low in rotary motion for the G0.5 emulsion. The test was conducted with oil also for comparison (Figure \ref{fig:6}f). Oil, with no internal physical structure, largely retraced its CoF trend, although slightly higher values appeared at very low loads (1–2 N). In contrast, the emulsion closely retraced the CoF path during unloading, showing only minor hysteresis. This behavior indicates the reversible assembly of the clay–droplet network governed by thixotropy, enabling the Pickering emulsion to recover its lubricating state and demonstrating its potential as a responsive semi-solid lubricant.

To isolate the effect of Garamite, suspensions with varying Garamite concentrations in water and oil were tested (Figure S6:SI). In water, G0.5 and G3 exhibited a comparable CoF to that of pure water. Similarly, in oil, low concentrations (G0.5 and G3) exhibited friction behavior comparable to that of the base oil. A reduction in friction was observed only at higher concentration (G10), although the improvement remained less pronounced than that achieved with the G0.5 emulsion. These results indicate that particle addition alone is insufficient for effective lubrication, and high loadings are economically undesirable.

\subsection{Wear Performance}\label{subsec2.4}

\begin{figure*}[tb]
    \centering     
    \includegraphics[width=\linewidth]{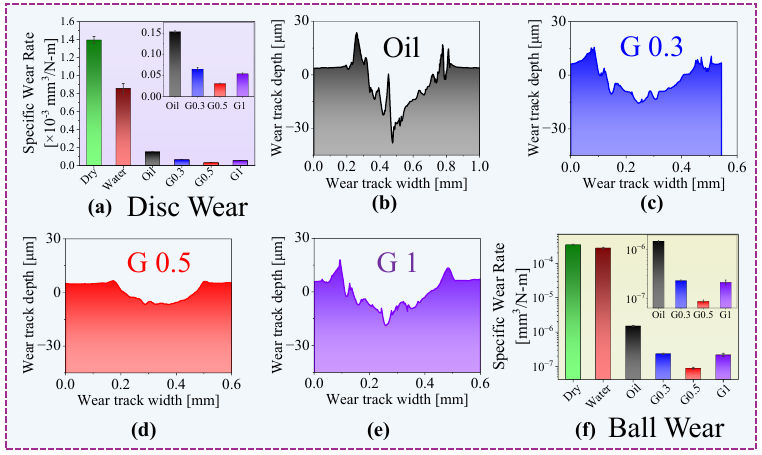}
    \caption{(a) Specific wear rate of the disc. Representative cross-sectional profiles of the wear tracks lubricated with (b) base oil, (c) G0.3, (d) G0.5, and (e) G1. (f) Specific wear rate of the ball. The inset shows a magnified view of the wear rates for clarity. Error bars denote the standard deviation; where not visible, the deviation falls within the column size. The tests were performed at 90 N and 0.1 m s\textsuperscript{-1}.}
    \label{fig:7}
\end{figure*}

The specific wear rate (SWR) of the disc lubricated with oil and emulsions is shown in \textbf{Figure \ref{fig:7}a}, with dry and water lubrication included for comparison. The SWR results followed the same trend as the friction results presented in Figure \ref{fig:6}d. Overall, G0.5 reduced SWR by $\approx$ 80\% compared to oil and $\approx$ 96\% compared to water, demonstrating excellent anti-friction and anti-wear performance. Representative wear profiles of the discs are shown in Figure \ref{fig:7}b–e. The oil-lubricated surface exhibits a wide, deep, and rough wear track. In contrast, G0.3 and G1 exhibited milder wear, whereas G0.5 displayed a smooth and shallow track. This observation supports the earlier mechanism that clay particles cap surface asperities and smoothen the contact path in the optimally structured emulsion. The wear of the AISI 304 steel ball counterface was also analyzed (Figure \ref{fig:7}f). The ball tested with the G0.5 emulsion showed the lowest wear, consistent with the disc results, achieving an overall $\approx$ 94\% reduction in SWR compared to oil lubrication.

%\FloatBarrier

\subsubsection{Droplet morphology after wear}\label{subsubsec2.4.1}

\begin{figure*}[tb]
    \centering     
    \includegraphics[width=\linewidth]{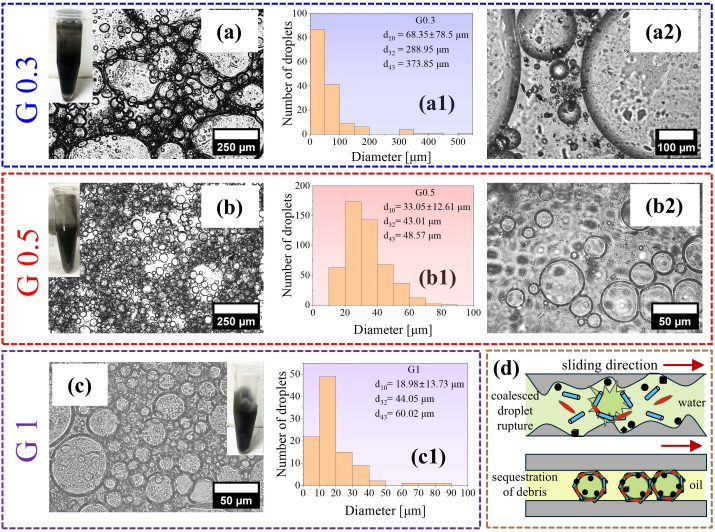}
    \caption{Optical micrographs and corresponding droplet size distribution histograms of emulsions after tribological testing: (a–a2) G0.3 and (b–b2) G0.5, where a2 and b2 show higher-magnification views of representative droplets; (c, c1) G1. (d) Schematic illustration of rupture of coalesced droplet and sequestration of debris by stable droplet. Tests were conducted at 90 N and 0.1 m s\textsuperscript{-1} for 10 min.}
    \label{fig:8}
\end{figure*}

\textbf{Figure \ref{fig:8}} presents optical micrographs of the Pickering emulsions following tribological testing. In the G0.3 system (Figure \ref{fig:8}a), the morphology is dominated by severe coalescence, forming massive, irregular droplets exceeding 500 $\mu\text{m}$ ($d_{43} = 373.85\ \mu\text{m}$, Figure \ref{fig:8}a1). This drastic increase from the fresh state confirms that the weak, aggregated Garamite shell cannot withstand the frictional shear. In contrast, the G0.5 system (Figure \ref{fig:8}b) demonstrated superior structural integrity. Despite the high shear forces, it maintained predominantly spherical droplets and a relatively narrow distribution ($d_{43} = 48.57\ \mu\text{m}$, Figure \ref{fig:8}b1). While minor flocculation is visible, the absence of coalescence correlates with the DSC findings of maximum interfacial confinement, providing a robust physical barrier against breakdown. In contrast, the G1 system (Figure \ref{fig:8}c) exhibited unexpected coarsening, with $d_{43}$ increasing to 60.02 μm (Figure \ref{fig:8}c1). This suggests that excess clay and localized aggregation, previously noted in the DSC thermograms, may facilitate droplet fusion under high shear.

High-magnification images (Figures \ref{fig:8}a2–b2) show the presence of steel wear debris (dark) within and outside the droplet boundaries. In the G0.3 system (Figure \ref{fig:8}a2), the rupture of coalesced droplets during sliding released both water and abrasive debris into the tribo-contact (Figure \ref{fig:8}d), synergistically increasing the friction and wear. In the G0.5 system (Figure \ref{fig:8}b2), wear debris was notably less prevalent. Although minor debris may be generated during the initial sliding phase, it is rapidly sequestered within stable droplets (Figure \ref{fig:8}d). These droplets act as ``micro-reservoirs," effectively capturing particles and preventing their recirculation in the contact zone, yielding the lowest observed friction and wear rates.

%\FloatBarrier

\subsubsection{Worn Surface Analysis}\label{subsubsec2.4.2}

\begin{figure*}[tb]
    \centering     
    \includegraphics[width=\linewidth]{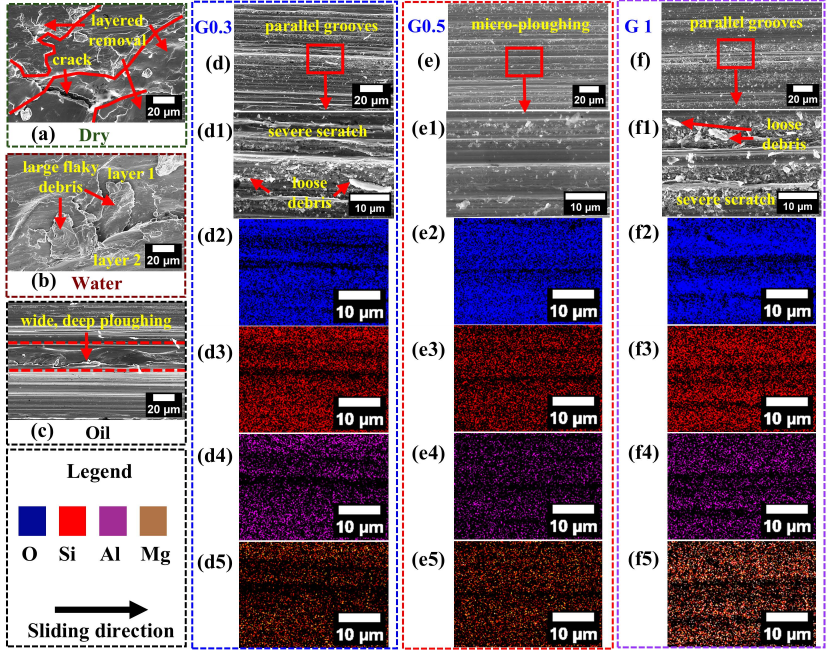}
    \caption{Post-tribological SEM micrographs of worn surfaces obtained under different lubrication conditions: (a) dry sliding, (b) water lubrication, and (c) oil. Worn surfaces lubricated with Garamite-containing emulsions are shown for (d) G0.3, (e) G0.5, and (f) G1, with the corresponding higher magnification views in (d1), (e1), and (f1), respectively. EDS maps for the respective wear tracks are presented in (d2–d5), (e2–e5), and (f2–f5).}
    \label{fig:9}
\end{figure*}

\textbf{Figure \ref{fig:9}} shows SEM micrographs and energy-dispersive X-ray spectroscopy (EDS) elemental maps of the worn surfaces. Under dry (Figure \ref{fig:9}a) and water lubrication (Figure \ref{fig:9}b), the surfaces exhibited severe damage. Dry sliding resulted in severe adhesive wear with layered delamination and deep cracks. Water lubrication produced massive flaky debris, indicating that its low viscosity and poor film-forming ability failed to separate the contacting surfaces. Oil lubrication (Figure \ref{fig:9}c) improved the condition but still showed wide and deep plowing tracks, suggesting that oil alone cannot sustain a sufficient boundary film at a high load.

The G0.3 system (Figure \ref{fig:9}d, d1) exhibited parallel grooves and severe scratches with loose debris. As observed in the post-test optical images (Figure \ref{fig:8}a2), the ineffective entrapment of debris from the steel surface promoted three-body-abrasion. The EDS maps (d2–d5) show scattered Si, Al, and Mg signals with a patchy distribution, indicating an unstable and discontinuous protective layer. In contrast, the G0.5 system (Figure \ref{fig:9}e, e1) exhibited the smoothest worn surface with only fine micro-plowing and minimal debris. This correlates with the stable post-test emulsion (Figure \ref{fig:8}b), in which the robust droplets effectively sequestered the wear debris. The EDS maps (e2–e5) show a dense and uniform distribution of Si, Al, and Mg along the wear track, indicating the formation of a cohesive clay-rich tribofilm. The surface lubricated with the G1 emulsion (Figure \ref{fig:9}f, f1) also exhibited scratches and loose debris, reflecting its limited protective capability (EDS maps: Figure \ref{fig:9} f2-f5). Overall, G0.5 effectively suppressed three-body abrasion, whereas the uniform clay-based tribofilm provided superior protection against metal-to-metal contact compared with oil alone.

%\FloatBarrier

%\subsubsection{Raman Analysis}

\subsubsection{Lubrication Mechanism}\label{subsubsec2.4.3}

\begin{figure*}[tb]
    \centering     
    \includegraphics[width=\linewidth]{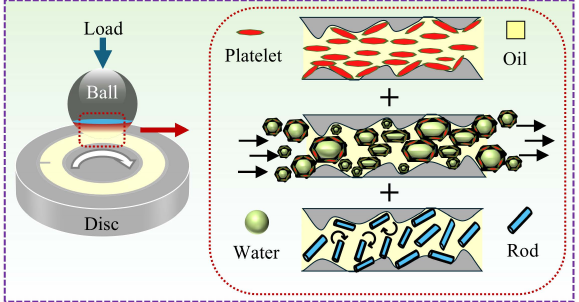}
    \caption{Schematic illustration of the lubrication mechanism of Garamite stabilised Pickering emulsion}
    \label{fig:10}
\end{figure*}
%\FloatBarrier

Based on the friction and wear analyses, a lubrication mechanism for the nanoclay-stabilized Pickering emulsion is proposed (\textbf{Figure \ref{fig:10}}). The superior tribological performance arises from the combined effects of thixotropy, anisotropic nanoclay morphology, and the strong armoring of water droplets. These effects together form a multiphase interfacial film that adapts to shear and load in the boundary-to-mixed regimes. At the asperities and valleys, the oil forms an adsorbed boundary film with low shear strength, while residual nanoclay accumulates to generate a compact clay-rich tribo-film as observed in Figure \ref{fig:9}. Within this layer, aligned platelets enable lamellar sliding, and rod-like particles provide rolling and load-bearing effects, thereby reducing the shear resistance and minimizing direct asperity contact.

Simultaneously, nanoclay-armored water droplets act as deformable micro-reservoirs that enter contact in a near-spherical shape and elastically deform under load, providing localized cushioning and reducing the real area of contact. The droplets recover their shape upon exiting the shear zone, indicating high structural stability and resistance to rupture, as observed in Figure \ref{fig:8}b2. The thixotropic nature of the emulsion enables shear thinning for efficient entrainment, followed by rapid structural recovery to maintain the droplet network and prevent lubricant starvation. Although all emulsions outperformed the base oil, G0.5 exhibited optimal performance owing to a balanced combination of droplet stability, particle availability, and rheological response. In contrast, in G0.3 and G1, reduced stability leads to droplet rupture under high stress; the released water, owing to its low pressure–viscosity coefficient, fails to sustain asperity separation, resulting in increased contact and wear. Consequently, the G0.5 emulsion formed a dense, load-bearing interfacial layer that delivered superior friction and wear resistance under severe conditions.

\section{Conclusions}\label{sec3}

This study demonstrated the feasibility of using Garamite nano-organoclay as an effective stabilizer for W/O Pickering emulsions. It also established its potential as an eco-conscious thixotropic lubricant for metallic tribosystems under very high pressure. The rheological and tribological responses were systematically tuned by varying the clay concentration (0.3, 0.5 and 1 wt.\%). The key findings are as follows:
 
\begin{enumerate}
    \item Increasing Garamite content significantly enhanced thixotropy, with higher dynamic moduli ($G'$, $G''$) and pronounced shear-thinning behavior, attributed to the formation of a strengthened excess particle bridged network around water droplets in the continuous oil phase.

    \item The G0.5 emulsion showed the best tribological performance, with $\approx$41\% and $\approx$84\% reduction in CoF, and $\approx$80\% and $\approx$96\% reduction in SWR, compared to oil and water, respectively.
    
    \item The emulsion exhibited lubrication performance sensitive to the sliding direction. It also demonstrated memory-dependent lubrication performance, indicating the adaptive assembly of the clay–droplet network governed by thixotropy.
    
    \item The superior performance originated from the synergistic interplay of thixotropy, anisotropic nanoclay morphology, and stable droplet armoring, which together formed a robust, adaptive multiphase interfacial film.
\end{enumerate}

Overall, the tribological behavior was governed by both the rheology and hierarchical microstructure of the nanoclay, arising from its platelet-rod morphology. These findings provide new insights into the structure-property relationships in Pickering emulsions and establish a pathway for designing next-generation sustainable metal-working fluids for advanced engineering applications.

\section{Experimental Section}\label{sec4}
\subsection{Materials and Chemicals}\label{subsec4.1}

Organophilic phyllosilicate nanoclay (Garamite-1958\textsuperscript{\textregistered}) was procured from BYK Additives Inc. It is a hybrid sepiolite-montmorillonite organoclay modified with alkyl quaternary ammonium ions \cite{Schubel2006CharacterisationReinforcement,Miller2011TWOCOMPOSITIONS}. It possesses high bulk density, unique platelet–rod morphology \cite{Liu2026Garamite-reinforcedDistillation} and provides effective sag resistance and anti-settling performance \cite{Miller2011TWOCOMPOSITIONS}. 

W/O Pickering emulsions were prepared using sunflower oil (Fortune Sunlite, locally procured) and Millipore\textsuperscript{\textregistered} water (resistivity 18.2 M$\Omega$-cm). Tribological tests were performed on AISI 304 austenitic stainless-steel discs with a hardness of 25 HRC. Circular discs ($\phi$50$\times$3 mm) were wire-cut using electro-discharge machining. The surface preparation involved sequential polishing with SiC papers from P320 to P5000 grit, followed by diamond polishing (6, 3, and 1 \textmu m). The samples were ultrasonically cleaned in acetone and deionized water, blow-dried, and finally wiped with acetone before testing to remove any residual contaminants.

\subsection{Preparation of Emulsion}\label{subsec4.2}

Emulsions were prepared by dispersing Garamite in sunflower oil using a magnetic stirrer (IKA\textsuperscript{\textregistered} C-MAG HS 7) at high speed and $25 ^{\circ}$C. After complete addition, the suspension was stirred for 10 min. to ensure uniform dispersion. Water was then added dropwise under continuous stirring, followed by an additional 10 min. of mixing to obtain a homogeneous emulsion. The mixture was then gently mixed manually for a few minutes to remove any local inhomogeneities. Table \ref{table:1} shows the details of the initially prepared emulsions with different water-oil weight ratios. 

\begin{table}[h!]
\centering
\caption{Compositions of the emulsions investigated}
\resizebox{\linewidth}{!}{%
\label{table:1}
\begin{tabular}{c c c c}
\hline
\textbf{Label} & \textbf{Water to oil ratio (by wt.)} & \textbf{Volume fraction of water ($\phi$)} & \textbf{Garamite (wt.\%)} \\
\hline
W8O2 & 4:1 & 0.78 & 0.5 \\
W7O3 & 7:3 & 0.68 & 0.5 \\
W6O4 & 3:2 & 0.58 & 0.5 \\
W5O5 & 1:1 & 0.48 & 0.5 \\
W4O6 & 2:3 & 0.38 & 0.5 \\
\hline
\end{tabular}}

\vspace{1mm}
{\footnotesize W8O2: maximum water content; W6O4: primary system investigated; W4O6: minimum water content; $\rho_{water}$= 0.997 g/cm\textsuperscript{3} and $\rho_{oil}$= 0.92 g/cm\textsuperscript{3} at $20 ^{\circ}$C.}
\end{table}

 \subsection{Characterization of Nanoclay and Emulsion}\label{subsec4.3}

The morphology of the Garamite nanoclay was examined using field-emission SEM (FESEM, NOVA NANOSEM 450). Phase identification was performed via XRD using a Rigaku MiniFlex-600 diffractometer with Cu K$\alpha$ radiation ($\lambda$ = 1.5405 \AA). Data were collected over a 2$\theta$ range of 5–100$^{\circ}$ at a scan rate of $2^{\circ}$ min\textsuperscript{-1} and a step size of $0.015^{\circ}$, with the X-ray source operated at 30 kV and 10 mA. SAXS measurements were performed on Garamite powder using a SAXSpoint~5.0 instrument (Anton Paar, Graz, Austria) equipped with a Cu~K$\alpha$ microfocus source ($\lambda = 0.154$~nm), scatterless pinhole collimation, and an EIGER2~R photon-counting detector. The measurements were performed under vacuum at room temperature. A sample-to-detector distance of 1599~mm provided a $q$ range of $\approx$ 0.02--2~nm$^{-1}$ and was calibrated using silver behenate. For each sample, five frames were collected with an exposure time of 300 s and averaged. The data were processed using \textit{SAXSanalysis} software. The \textit{SASview} software was used for the analysis. The elemental composition with corresponding weight percentages were determined using a wavelength-dispersive XRF spectrometer (Rigaku ZSX Primus II) in accordance with ASTM E1621–22 \cite{ASTME1621222022StandardSpectrometry}.

FTIR spectra of the clay and emulsions were recorded using a Perkin Elmer Spectrum 2 spectrometer over 4000–400 cm\textsuperscript{-1}. Spectra were obtained by averaging 20 scans at a resolution of 4 cm\textsuperscript{-1} using Happ–Genzel apodization. The reference spectrum of KBr was subtracted to obtain the final clay spectrum. All measurements were performed at room temperature. 

Thermal analysis in the form of DSC was performed using a Discovery DSC 25 (TA Instruments) in a dry nitrogen atmosphere. Samples were hermetically sealed in a standard aluminum pan and subjected to a protocol consisting of an initial equilibration at $-60 ^{\circ}$C for 1 min, followed by heating to $80 ^{\circ}$C at $1.5 ^{\circ}$C/min, isothermal hold for 1 min, cooling to $-60 ^{\circ}$C at $1.5 ^{\circ}$C/min, and final isothermal equilibration (1 min) with an empty pan as a reference. The heat flow was measured as a function of the temperature. The crystallization ($T_c$) and fusion ($T_f$) temperatures were determined from the peak maxima, and the corresponding enthalpies ($\Delta H$) were calculated by numerical integration of the exothermic and endothermic transitions.

The morphology and distribution of the water droplets in the as-prepared emulsions were examined using a polarized optical microscope (Leica DM2700P) 12 h after sample preparation. The droplet size distribution was determined by analyzing  the droplets using the ImageJ software. The droplet morphology and distribution after tribological testing were evaluated using optical microscopy.

Microstructural images for measuring the thickness of the nanoclay-armored shell around the water droplets were obtained using a polarized optical microscope (Leica DM2700P, Germany). Nile Red was used as a fluorescent dye, with excitation and emission spectra of 538$\pm$45 nm and above 590 nm, respectively. To stain the Garamite particles, 1 $\mu$L of a 1 wt.\% Nile Red solution in acetone was added to 10 mL of the oil phase containing dispersed Garamite and gently stirred for 1 h at $25 ^{\circ}$C. The required amount of water was then added to form an emulsion.

The storage stability of the emulsions was evaluated using CI over 30 days. After emulsion preparation, 25 mL of each sample was stored in tightly sealed centrifuge tubes and kept undisturbed at $25 ^{\circ}$C. The creaming index is defined as the ratio of the oil layer height ($H_s$) to the total emulsion height ($H_E$) and is calculated as follows:

\begin{equation}
    CI (\%)= (H_S/H_E)\times100
\end{equation}

The electrical conductivity was measured at room temperature using a  HORIBA Scientific conductivity meter (LAQUA PC1100, HORIBA, Japan) to monitor the emulsion stability and phase inversion over 30 days.

The interfacial surface tension of the emulsion (oil continuous phase) in water was measured using a Kr$\ddot{u}$ss DSA 25E Goniometer with a 0.125 mm (outer diameter) syringe. A droplet of approximately 5$\mu L$ is generated carefully inside the cuvette filled with Millipore water for interfacial tension measurement.

\subsection{Rheological Tests}\label{subsec4.4}

The rheological experiments were carried out on a stress-controlled Anton Paar MCR-501 Rheometer at a temperature of $25  ^{\circ}$C using a sandblasted (to avoid wall slip) concentric cylinder geometry (cup and bob) with bob diameter and length as 26.650 mm and  40.020 mm, respectively. The inner diameter of the cup was 28.915 mm. A solvent trap was employed to minimize evaporation loss during the experiment. To erase the previous microstructure build-up history of the sample and ensure that all experiments began at the same initial state, pre-shearing was performed at a large oscillatory strain amplitude of 4000\% with a frequency of 0.05 Hz for 10 min. After pre-shearing, the material was allowed to age for a waiting time ($t_w$) of 10 min while maintaining the linear regime by applying a small oscillatory strain amplitude of 0.1\% and a frequency of 0.1 Hz. These two steps were followed before starting all the rheological experiments. 

To understand the microstructural evolution in the system or physical aging behavior, a time sweep experiment was performed for 1 h at a strain amplitude of 0.1\% and frequency of 0.1 Hz immediately after the pre-shearing step to study the evolution of $G'$ (storage modulus) and $G''$ (loss modulus) with waiting time ($t_w$). Furthermore, to examine the material's viscoelastic response ($G'$, $G''$) variation with the observation time scale, frequency sweep were performed from high to low frequencies (19.4–0.01 Hz), well within the linear regime  ($\gamma_0 = 0.1 \% $). 

To understand the yielding behavior, dynamic oscillatory strain sweeps were performed from ($\gamma_0 = 0.01 - 25.1\% $) at a frequency of 0.1 Hz. Finally, the system was subjected to a steady-state shear rate sweep progressing from high to low shear rate (100-0.0001 $s^{-1}$) to determine the flow curve and characterize the flow behavior. The upper limits of the explored shear rate ($\dot\gamma$), strain amplitude ($\gamma$), and frequency in their respective sweep experiments were limited by the onset of turbulence and significant nonlinearity in the system.

\subsection{Tribological Tests}\label{subsec4.5}

Rtec MFT-5000 and MFT-2000 tribometers (Figure S7: SI) were used for tribology tests. The balls used as counterface had a diameter of 6 mm, and the maximum average surface roughness ($R_a$) was 0.20 \textmu m.

For comprehensive frictional performance, the friction coefficients of the as-prepared emulsions were measured as functions of entrainment speed and normal load. The sliding speed was varied from 0.005 to 2 m s\textsuperscript{-1} at a constant load of 5 N. The load was varied from 1 to 120 N at a constant speed of 0.1 m s\textsuperscript{-1}. This approach enabled the evaluation of the lubrication behavior across different lubrication regimes. The speed and load were increased stepwise, with each step lasting 7 min. to ensure stable measurement. To assess the directional effects, reciprocating tests were conducted using the same load range as rotary tests. A stroke length of 8 mm and frequency of 6.2 Hz were applied, yielding an average linear sliding speed of approximately 0.1 m s\textsuperscript{-1}. Based on the friction trends under varying loads, the wear and load-bearing performance were further evaluated using unidirectional rotary tests under submerged lubrication at 90 N and 0.1 m s\textsuperscript{-1} for 10 min. The tribological performance was compared with that of dry, water, and oil lubrications. 

All tests followed ASTM G99 \cite{ASTMInternationalG99232023TestApparatus} for rotary motion and ASTM G133 (Procedure A) \cite{ASTMG133-222022TestWear} for reciprocating motion, with permitted modifications to the test parameters. The experiments were conducted at $23\pm0.3^{\circ}$C and 57$\pm$3\% relative humidity. Approximately 5 g of the emulsion was uniformly spread on the disc surface for each test. Prior to each test, the friction force sensors of both tribometers were checked using known static weights to ensure accurate measurements. A normal load was applied with a maximum deviation of $\pm$2.0\% and monitored using a high-precision 2D capacitance-based load cell with minimal drift \cite{Kumar2026SynergisticTemperature}. The electroservo drives ensured stable load control throughout the tests. The initial test conditions were carefully maintained to ensure consistency across all experiments. The data were acquired at a sampling rate of 1 kHz and averaged over 10 points. The friction coefficients are reported as the mean and standard deviation from at least three measurements.

\subsection{Basic Surface Analysis}\label{subsec4.6}

The surface roughness and wear profiles of the AISI 304 discs were measured using a non-contact 3D optical profilometer with white light interferometry (Contour GT-K, Bruker). The wear volume was calculated to determine the SWR (mm\textsuperscript{3}/N-m) using the ISO 18535:2016 method, as described in \cite{Kumar2025Nano-silicaLubricant,Kumar2024DrySubstrates,Kumar2024TribologicalNanoparticles}. The wear scars on the AISI 304 balls were examined using an optical microscope (Olympus BX51M). The worn disc surfaces were further analyzed using FESEM, and the elemental composition was determined using EDS (Oxford Instruments) with INCA and Aztec software.

\section*{CRediT authorship contribution statement}
\textbf{AK}: Methodology, Investigation, Data curation, Formal analysis, Writing-original draft. \textbf{RY}: Methodology, Investigation, Data curation, Formal analysis, Writing–original draft. \textbf{YMJ}: Conceptualization, Funding acquisition, Supervision, Writing–review and editing. \textbf{MKS}: Conceptualization, Funding acquisition, Supervision, Writing–review and editing. 

\section*{Declarations}
The authors declare no competing financial interest.

\section*{Acknowledgement}
YMJ acknowledges financial support from the Science and Engineering Research Board, Government of India (Grant numbers: CRG/2022/004868 and JCB/2022/000040).

\section*{Supporting Information}
Supporting information is available. 

\section*{Data availability}
Data will be made available on reasonable request.

% \appendix*
% \input{sections/appendix1.tex}

%\bibliography{References}

\begin{thebibliography}{74}%
	\makeatletter
	\providecommand \@ifxundefined [1]{%
		\@ifx{#1\undefined}
	}%
	\providecommand \@ifnum [1]{%
		\ifnum #1\expandafter \@firstoftwo
		\else \expandafter \@secondoftwo
		\fi
	}%
	\providecommand \@ifx [1]{%
		\ifx #1\expandafter \@firstoftwo
		\else \expandafter \@secondoftwo
		\fi
	}%
	\providecommand \natexlab [1]{#1}%
	\providecommand \enquote  [1]{``#1''}%
	\providecommand \bibnamefont  [1]{#1}%
	\providecommand \bibfnamefont [1]{#1}%
	\providecommand \citenamefont [1]{#1}%
	\providecommand \href@noop [0]{\@secondoftwo}%
	\providecommand \href [0]{\begingroup \@sanitize@url \@href}%
	\providecommand \@href[1]{\@@startlink{#1}\@@href}%
	\providecommand \@@href[1]{\endgroup#1\@@endlink}%
	\providecommand \@sanitize@url [0]{\catcode `\\12\catcode `\$12\catcode
		`\&12\catcode `\#12\catcode `\^12\catcode `\_12\catcode `\%12\relax}%
	\providecommand \@@startlink[1]{}%
	\providecommand \@@endlink[0]{}%
	\providecommand \url  [0]{\begingroup\@sanitize@url \@url }%
	\providecommand \@url [1]{\endgroup\@href {#1}{\urlprefix }}%
	\providecommand \urlprefix  [0]{URL }%
	\providecommand \Eprint [0]{\href }%
	\providecommand \doibase [0]{http://dx.doi.org/}%
	\providecommand \selectlanguage [0]{\@gobble}%
	\providecommand \bibinfo  [0]{\@secondoftwo}%
	\providecommand \bibfield  [0]{\@secondoftwo}%
	\providecommand \translation [1]{[#1]}%
	\providecommand \BibitemOpen [0]{}%
	\providecommand \bibitemStop [0]{}%
	\providecommand \bibitemNoStop [0]{.\EOS\space}%
	\providecommand \EOS [0]{\spacefactor3000\relax}%
	\providecommand \BibitemShut  [1]{\csname bibitem#1\endcsname}%
	\let\auto@bib@innerbib\@empty
	%</preamble>
	\bibitem [{\citenamefont {Singh}(2022)}]{Singh2022AqueousLubrication}%
	\BibitemOpen
	\bibfield  {author} {\bibinfo {author} {\bibfnamefont {M.~K.}\ \bibnamefont
			{Singh}},\ }in\ \href {\doibase https://doi.org/10.1201/9781003092162} {\emph
		{\bibinfo {booktitle} {Tribology and Sustainability}}},\ Vol.~\bibinfo
	{volume} {1},\ \bibinfo {editor} {edited by\ \bibinfo {editor} {\bibfnamefont
			{J.~K.}\ \bibnamefont {Katiyar}}, \bibinfo {editor} {\bibfnamefont
			{M.}~\bibnamefont {Irfan Ul~Haq}}, \bibinfo {editor} {\bibfnamefont
			{A.}~\bibnamefont {Raina}}, \bibinfo {editor} {\bibfnamefont
			{S.}~\bibnamefont {Jayalakshmi}}, \ and\ \bibinfo {editor} {\bibfnamefont
			{R.~A.}\ \bibnamefont {Singh}}}\ (\bibinfo  {publisher} {CRC Press},\
	\bibinfo {address} {Boca Raton and Oxon},\ \bibinfo {year} {2022})\ \bibinfo
	{edition} {1st}\ ed.,\ Chap.~\bibinfo {chapter} {20}, pp.\ \bibinfo {pages}
	{397--412}\BibitemShut {NoStop}%
	\bibitem [{\citenamefont {Singh}\ \emph {et~al.}(2015)\citenamefont {Singh},
		\citenamefont {Ilg}, \citenamefont {Espinosa-Marzal}, \citenamefont
		{Kr{\"{o}}ger},\ and\ \citenamefont {Spencer}}]{Singh2015PolymerExperiments}%
	\BibitemOpen
	\bibfield  {author} {\bibinfo {author} {\bibfnamefont {M.~K.}\ \bibnamefont
			{Singh}}, \bibinfo {author} {\bibfnamefont {P.}~\bibnamefont {Ilg}}, \bibinfo
		{author} {\bibfnamefont {R.~M.}\ \bibnamefont {Espinosa-Marzal}}, \bibinfo
		{author} {\bibfnamefont {M.}~\bibnamefont {Kr{\"{o}}ger}}, \ and\ \bibinfo
		{author} {\bibfnamefont {N.~D.}\ \bibnamefont {Spencer}},\ }\href {\doibase
		10.1021/acs.langmuir.5b00641} {\bibfield  {journal} {\bibinfo  {journal}
			{Langmuir}\ }\textbf {\bibinfo {volume} {31}},\ \bibinfo {pages} {4798}
		(\bibinfo {year} {2015})}\BibitemShut {NoStop}%
	\bibitem [{\citenamefont {Chen}\ \emph {et~al.}(2022)\citenamefont {Chen},
		\citenamefont {Wang}, \citenamefont {Lu},\ and\ \citenamefont
		{Xu}}]{Chen2022Surfactant-ModifiedAnticorrosion}%
	\BibitemOpen
	\bibfield  {author} {\bibinfo {author} {\bibfnamefont {S.}~\bibnamefont
			{Chen}}, \bibinfo {author} {\bibfnamefont {J.}~\bibnamefont {Wang}}, \bibinfo
		{author} {\bibfnamefont {H.}~\bibnamefont {Lu}}, \ and\ \bibinfo {author}
		{\bibfnamefont {L.}~\bibnamefont {Xu}},\ }\href {\doibase
		10.1021/acssuschemeng.2c01753} {\bibfield  {journal} {\bibinfo  {journal}
			{ACS Sustainable Chemistry and Engineering}\ }\textbf {\bibinfo {volume}
			{10}},\ \bibinfo {pages} {10816} (\bibinfo {year} {2022})}\BibitemShut
	{NoStop}%
	\bibitem [{\citenamefont {Bao}\ \emph {et~al.}(2022)\citenamefont {Bao},
		\citenamefont {Liu}, \citenamefont {Zheng}, \citenamefont {Yao},\ and\
		\citenamefont {Xu}}]{Bao2022AEmulsion}%
	\BibitemOpen
	\bibfield  {author} {\bibinfo {author} {\bibfnamefont {Y.}~\bibnamefont
			{Bao}}, \bibinfo {author} {\bibfnamefont {K.}~\bibnamefont {Liu}}, \bibinfo
		{author} {\bibfnamefont {Q.}~\bibnamefont {Zheng}}, \bibinfo {author}
		{\bibfnamefont {L.}~\bibnamefont {Yao}}, \ and\ \bibinfo {author}
		{\bibfnamefont {Y.}~\bibnamefont {Xu}},\ }\href {\doibase 10.1115/1.4052480}
	{\bibfield  {journal} {\bibinfo  {journal} {Journal of Tribology}\ }\textbf
		{\bibinfo {volume} {144}} (\bibinfo {year} {2022}),\
		10.1115/1.4052480}\BibitemShut {NoStop}%
	\bibitem [{\citenamefont {Liu}\ \emph {et~al.}(2024)\citenamefont {Liu},
		\citenamefont {Wang}, \citenamefont {Yang}, \citenamefont {Chen},
		\citenamefont {Yang},\ and\ \citenamefont
		{Zhang}}]{Liu2024TribologicalSurfactants}%
	\BibitemOpen
	\bibfield  {author} {\bibinfo {author} {\bibfnamefont {H.}~\bibnamefont
			{Liu}}, \bibinfo {author} {\bibfnamefont {X.}~\bibnamefont {Wang}}, \bibinfo
		{author} {\bibfnamefont {T.}~\bibnamefont {Yang}}, \bibinfo {author}
		{\bibfnamefont {S.}~\bibnamefont {Chen}}, \bibinfo {author} {\bibfnamefont
			{S.}~\bibnamefont {Yang}}, \ and\ \bibinfo {author} {\bibfnamefont
			{X.}~\bibnamefont {Zhang}},\ }\href {\doibase 10.1016/j.colsurfa.2024.134206}
	{\bibfield  {journal} {\bibinfo  {journal} {Colloids and Surfaces A:
				Physicochemical and Engineering Aspects}\ }\textbf {\bibinfo {volume}
			{695}},\ \bibinfo {pages} {134206} (\bibinfo {year} {2024})}\BibitemShut
	{NoStop}%
	\bibitem [{\citenamefont {Du}\ \emph {et~al.}(2025)\citenamefont {Du},
		\citenamefont {Zhou}, \citenamefont {Huang},\ and\ \citenamefont
		{Meng}}]{Du2025InvestigationAnalogues}%
	\BibitemOpen
	\bibfield  {author} {\bibinfo {author} {\bibfnamefont {L.}~\bibnamefont
			{Du}}, \bibinfo {author} {\bibfnamefont {S.}~\bibnamefont {Zhou}}, \bibinfo
		{author} {\bibfnamefont {Y.}~\bibnamefont {Huang}}, \ and\ \bibinfo {author}
		{\bibfnamefont {Z.}~\bibnamefont {Meng}},\ }\href {\doibase
		10.1016/j.foodchem.2024.142121} {\bibfield  {journal} {\bibinfo  {journal}
			{Food Chemistry}\ }\textbf {\bibinfo {volume} {465}},\ \bibinfo {pages}
		{142121} (\bibinfo {year} {2025})}\BibitemShut {NoStop}%
	\bibitem [{\citenamefont {Feng}\ \emph {et~al.}(2025)\citenamefont {Feng},
		\citenamefont {Ding}, \citenamefont {Zhu}, \citenamefont {Wang},
		\citenamefont {Zhao}, \citenamefont {Yang}, \citenamefont {Wang},
		\citenamefont {Xia},\ and\ \citenamefont
		{Huang}}]{Feng2025Cinnamaldehyde-loadedPerspectives}%
	\BibitemOpen
	\bibfield  {author} {\bibinfo {author} {\bibfnamefont {T.}~\bibnamefont
			{Feng}}, \bibinfo {author} {\bibfnamefont {Q.}~\bibnamefont {Ding}}, \bibinfo
		{author} {\bibfnamefont {K.}~\bibnamefont {Zhu}}, \bibinfo {author}
		{\bibfnamefont {Y.}~\bibnamefont {Wang}}, \bibinfo {author} {\bibfnamefont
			{Q.}~\bibnamefont {Zhao}}, \bibinfo {author} {\bibfnamefont {F.}~\bibnamefont
			{Yang}}, \bibinfo {author} {\bibfnamefont {X.}~\bibnamefont {Wang}}, \bibinfo
		{author} {\bibfnamefont {S.}~\bibnamefont {Xia}}, \ and\ \bibinfo {author}
		{\bibfnamefont {Q.}~\bibnamefont {Huang}},\ }\href {\doibase
		10.1002/jsfa.70053} {\bibfield  {journal} {\bibinfo  {journal} {Journal of
				the Science of Food and Agriculture}\ }\textbf {\bibinfo {volume} {105}},\
		\bibinfo {pages} {8048} (\bibinfo {year} {2025})}\BibitemShut {NoStop}%
	\bibitem [{\citenamefont {Du}\ and\ \citenamefont
		{Meng}(2025)}]{Du2025WhippedCrystals}%
	\BibitemOpen
	\bibfield  {author} {\bibinfo {author} {\bibfnamefont {L.}~\bibnamefont
			{Du}}\ and\ \bibinfo {author} {\bibfnamefont {Z.}~\bibnamefont {Meng}},\
	}\href {\doibase 10.1016/j.foodhyd.2025.111050} {\bibfield  {journal}
		{\bibinfo  {journal} {Food Hydrocolloids}\ }\textbf {\bibinfo {volume}
			{163}},\ \bibinfo {pages} {111050} (\bibinfo {year} {2025})}\BibitemShut
	{NoStop}%
	\bibitem [{\citenamefont {Lu}\ \emph {et~al.}(2021{\natexlab{a}})\citenamefont
		{Lu}, \citenamefont {Zhou}, \citenamefont {Ye}, \citenamefont {Zhou},
		\citenamefont {Gao}, \citenamefont {Qin},\ and\ \citenamefont
		{Zhao}}]{Lu2021AParticles}%
	\BibitemOpen
	\bibfield  {author} {\bibinfo {author} {\bibfnamefont {Z.}~\bibnamefont
			{Lu}}, \bibinfo {author} {\bibfnamefont {S.}~\bibnamefont {Zhou}}, \bibinfo
		{author} {\bibfnamefont {F.}~\bibnamefont {Ye}}, \bibinfo {author}
		{\bibfnamefont {G.}~\bibnamefont {Zhou}}, \bibinfo {author} {\bibfnamefont
			{R.}~\bibnamefont {Gao}}, \bibinfo {author} {\bibfnamefont {D.}~\bibnamefont
			{Qin}}, \ and\ \bibinfo {author} {\bibfnamefont {G.}~\bibnamefont {Zhao}},\
	}\href {\doibase 10.1016/j.foodchem.2021.129418} {\bibfield  {journal}
		{\bibinfo  {journal} {Food Chemistry}\ }\textbf {\bibinfo {volume} {353}},\
		\bibinfo {pages} {129418} (\bibinfo {year} {2021}{\natexlab{a}})}\BibitemShut
	{NoStop}%
	\bibitem [{\citenamefont {Guo}\ \emph {et~al.}(2025)\citenamefont {Guo},
		\citenamefont {Wu}, \citenamefont {Guo}, \citenamefont {Hu}, \citenamefont
		{Ou}, \citenamefont {Zhu}, \citenamefont {Luo}, \citenamefont {Song},
		\citenamefont {He}, \citenamefont {He}, \citenamefont {Xu}, \citenamefont
		{Tang}, \citenamefont {Qin}, \citenamefont {Wang}, \citenamefont {Du},\ and\
		\citenamefont {Sun}}]{Guo2025Metalloparticle-EngineeredPathogens}%
	\BibitemOpen
	\bibfield  {author} {\bibinfo {author} {\bibfnamefont {Z.}~\bibnamefont
			{Guo}}, \bibinfo {author} {\bibfnamefont {F.}~\bibnamefont {Wu}}, \bibinfo
		{author} {\bibfnamefont {C.}~\bibnamefont {Guo}}, \bibinfo {author}
		{\bibfnamefont {R.}~\bibnamefont {Hu}}, \bibinfo {author} {\bibfnamefont
			{Y.}~\bibnamefont {Ou}}, \bibinfo {author} {\bibfnamefont {Y.}~\bibnamefont
			{Zhu}}, \bibinfo {author} {\bibfnamefont {S.}~\bibnamefont {Luo}}, \bibinfo
		{author} {\bibfnamefont {Y.}~\bibnamefont {Song}}, \bibinfo {author}
		{\bibfnamefont {P.}~\bibnamefont {He}}, \bibinfo {author} {\bibfnamefont
			{C.}~\bibnamefont {He}}, \bibinfo {author} {\bibfnamefont {Y.}~\bibnamefont
			{Xu}}, \bibinfo {author} {\bibfnamefont {X.}~\bibnamefont {Tang}}, \bibinfo
		{author} {\bibfnamefont {M.}~\bibnamefont {Qin}}, \bibinfo {author}
		{\bibfnamefont {H.}~\bibnamefont {Wang}}, \bibinfo {author} {\bibfnamefont
			{G.}~\bibnamefont {Du}}, \ and\ \bibinfo {author} {\bibfnamefont
			{X.}~\bibnamefont {Sun}},\ }\href {\doibase 10.1002/adma.202412627}
	{\bibfield  {journal} {\bibinfo  {journal} {Advanced Materials}\ }\textbf
		{\bibinfo {volume} {37}},\ \bibinfo {pages} {2412627} (\bibinfo {year}
		{2025})}\BibitemShut {NoStop}%
	\bibitem [{\citenamefont {Dai}\ \emph {et~al.}(2023)\citenamefont {Dai},
		\citenamefont {Zhang}, \citenamefont {Bao}, \citenamefont {Guo},
		\citenamefont {Jin}, \citenamefont {Yang}, \citenamefont {Zhang},
		\citenamefont {Liu}, \citenamefont {Gao}, \citenamefont {Ye}, \citenamefont
		{Wu}, \citenamefont {Liu}, \citenamefont {Zhao}, \citenamefont {Sheng},
		\citenamefont {Ren}, \citenamefont {Li}, \citenamefont {Fang}, \citenamefont
		{Wu}, \citenamefont {Ruan}, \citenamefont {Gu}, \citenamefont {Chen},\ and\
		\citenamefont {Zhao}}]{Dai2023InductionGel}%
	\BibitemOpen
	\bibfield  {author} {\bibinfo {author} {\bibfnamefont {X.}~\bibnamefont
			{Dai}}, \bibinfo {author} {\bibfnamefont {J.}~\bibnamefont {Zhang}}, \bibinfo
		{author} {\bibfnamefont {X.}~\bibnamefont {Bao}}, \bibinfo {author}
		{\bibfnamefont {Y.}~\bibnamefont {Guo}}, \bibinfo {author} {\bibfnamefont
			{Y.}~\bibnamefont {Jin}}, \bibinfo {author} {\bibfnamefont {C.}~\bibnamefont
			{Yang}}, \bibinfo {author} {\bibfnamefont {H.}~\bibnamefont {Zhang}},
		\bibinfo {author} {\bibfnamefont {L.}~\bibnamefont {Liu}}, \bibinfo {author}
		{\bibfnamefont {Y.}~\bibnamefont {Gao}}, \bibinfo {author} {\bibfnamefont
			{C.}~\bibnamefont {Ye}}, \bibinfo {author} {\bibfnamefont {W.}~\bibnamefont
			{Wu}}, \bibinfo {author} {\bibfnamefont {C.}~\bibnamefont {Liu}}, \bibinfo
		{author} {\bibfnamefont {C.~X.}\ \bibnamefont {Zhao}}, \bibinfo {author}
		{\bibfnamefont {J.}~\bibnamefont {Sheng}}, \bibinfo {author} {\bibfnamefont
			{E.}~\bibnamefont {Ren}}, \bibinfo {author} {\bibfnamefont {H.}~\bibnamefont
			{Li}}, \bibinfo {author} {\bibfnamefont {W.}~\bibnamefont {Fang}}, \bibinfo
		{author} {\bibfnamefont {B.}~\bibnamefont {Wu}}, \bibinfo {author}
		{\bibfnamefont {J.}~\bibnamefont {Ruan}}, \bibinfo {author} {\bibfnamefont
			{Z.}~\bibnamefont {Gu}}, \bibinfo {author} {\bibfnamefont {D.}~\bibnamefont
			{Chen}}, \ and\ \bibinfo {author} {\bibfnamefont {P.}~\bibnamefont {Zhao}},\
	}\href {\doibase 10.1002/adma.202303542} {\bibfield  {journal} {\bibinfo
			{journal} {Advanced Materials}\ }\textbf {\bibinfo {volume} {35}},\ \bibinfo
		{pages} {2303542} (\bibinfo {year} {2023})}\BibitemShut {NoStop}%
	\bibitem [{\citenamefont {Xia}\ \emph {et~al.}(2025)\citenamefont {Xia},
		\citenamefont {Li}, \citenamefont {Huang}, \citenamefont {Wang},
		\citenamefont {Xiong}, \citenamefont {Zhou}, \citenamefont {Li},
		\citenamefont {Lin}, \citenamefont {Tang},\ and\ \citenamefont
		{Zhang}}]{Xia2025AChemoembolization}%
	\BibitemOpen
	\bibfield  {author} {\bibinfo {author} {\bibfnamefont {X.}~\bibnamefont
			{Xia}}, \bibinfo {author} {\bibfnamefont {Y.}~\bibnamefont {Li}}, \bibinfo
		{author} {\bibfnamefont {R.}~\bibnamefont {Huang}}, \bibinfo {author}
		{\bibfnamefont {Y.}~\bibnamefont {Wang}}, \bibinfo {author} {\bibfnamefont
			{W.}~\bibnamefont {Xiong}}, \bibinfo {author} {\bibfnamefont
			{H.}~\bibnamefont {Zhou}}, \bibinfo {author} {\bibfnamefont {M.}~\bibnamefont
			{Li}}, \bibinfo {author} {\bibfnamefont {X.}~\bibnamefont {Lin}}, \bibinfo
		{author} {\bibfnamefont {Y.}~\bibnamefont {Tang}}, \ and\ \bibinfo {author}
		{\bibfnamefont {B.}~\bibnamefont {Zhang}},\ }\href {\doibase
		10.1002/advs.202410873} {\bibfield  {journal} {\bibinfo  {journal} {Advanced
				Science}\ }\textbf {\bibinfo {volume} {12}},\ \bibinfo {pages} {2410873}
		(\bibinfo {year} {2025})}\BibitemShut {NoStop}%
	\bibitem [{\citenamefont {Zhang}\ \emph
		{et~al.}(2025{\natexlab{a}})\citenamefont {Zhang}, \citenamefont {Li},
		\citenamefont {Carland}, \citenamefont {Ye}, \citenamefont {Bell},\ and\
		\citenamefont {Xu}}]{Zhang2025UntyingBio-Media}%
	\BibitemOpen
	\bibfield  {author} {\bibinfo {author} {\bibfnamefont {Y.}~\bibnamefont
			{Zhang}}, \bibinfo {author} {\bibfnamefont {C.}~\bibnamefont {Li}}, \bibinfo
		{author} {\bibfnamefont {R.}~\bibnamefont {Carland}}, \bibinfo {author}
		{\bibfnamefont {Z.}~\bibnamefont {Ye}}, \bibinfo {author} {\bibfnamefont
			{S.~E.}\ \bibnamefont {Bell}}, \ and\ \bibinfo {author} {\bibfnamefont
			{Y.}~\bibnamefont {Xu}},\ }\href {\doibase 10.1002/advs.202505714} {\bibfield
		{journal} {\bibinfo  {journal} {Advanced Science}\ }\textbf {\bibinfo
			{volume} {12}} (\bibinfo {year} {2025}{\natexlab{a}}),\
		10.1002/advs.202505714}\BibitemShut {NoStop}%
	\bibitem [{\citenamefont {Guan}\ \emph {et~al.}(2026)\citenamefont {Guan},
		\citenamefont {Ming}, \citenamefont {Zhou}, \citenamefont {Chio},
		\citenamefont {Jiang}, \citenamefont {Li}, \citenamefont {Tse}, \citenamefont
		{Xia},\ and\ \citenamefont {Ngai}}]{Guan2026EngineeringCarriers}%
	\BibitemOpen
	\bibfield  {author} {\bibinfo {author} {\bibfnamefont {X.}~\bibnamefont
			{Guan}}, \bibinfo {author} {\bibfnamefont {Y.}~\bibnamefont {Ming}}, \bibinfo
		{author} {\bibfnamefont {Y.}~\bibnamefont {Zhou}}, \bibinfo {author}
		{\bibfnamefont {C.~C.}\ \bibnamefont {Chio}}, \bibinfo {author}
		{\bibfnamefont {H.}~\bibnamefont {Jiang}}, \bibinfo {author} {\bibfnamefont
			{L.}~\bibnamefont {Li}}, \bibinfo {author} {\bibfnamefont {Y.~L.~S.}\
			\bibnamefont {Tse}}, \bibinfo {author} {\bibfnamefont {Y.}~\bibnamefont
			{Xia}}, \ and\ \bibinfo {author} {\bibfnamefont {T.}~\bibnamefont {Ngai}},\
	}\href {\doibase 10.1016/j.jcis.2025.138790} {\bibfield  {journal} {\bibinfo
			{journal} {Journal of Colloid and Interface Science}\ }\textbf {\bibinfo
			{volume} {702}},\ \bibinfo {pages} {138790} (\bibinfo {year}
		{2026})}\BibitemShut {NoStop}%
	\bibitem [{\citenamefont {Yao}\ \emph {et~al.}(2025)\citenamefont {Yao},
		\citenamefont {Feng}, \citenamefont {Ao}, \citenamefont {Zhang},
		\citenamefont {Zhang}, \citenamefont {Wang}, \citenamefont {Liu},
		\citenamefont {Wang},\ and\ \citenamefont {Yu}}]{Yao2025NaturalBacteria}%
	\BibitemOpen
	\bibfield  {author} {\bibinfo {author} {\bibfnamefont {Y.}~\bibnamefont
			{Yao}}, \bibinfo {author} {\bibfnamefont {J.}~\bibnamefont {Feng}}, \bibinfo
		{author} {\bibfnamefont {N.}~\bibnamefont {Ao}}, \bibinfo {author}
		{\bibfnamefont {Y.}~\bibnamefont {Zhang}}, \bibinfo {author} {\bibfnamefont
			{J.}~\bibnamefont {Zhang}}, \bibinfo {author} {\bibfnamefont
			{Y.}~\bibnamefont {Wang}}, \bibinfo {author} {\bibfnamefont {C.}~\bibnamefont
			{Liu}}, \bibinfo {author} {\bibfnamefont {M.}~\bibnamefont {Wang}}, \ and\
		\bibinfo {author} {\bibfnamefont {C.}~\bibnamefont {Yu}},\ }\href {\doibase
		10.1016/j.jcis.2024.09.066} {\bibfield  {journal} {\bibinfo  {journal}
			{Journal of Colloid and Interface Science}\ }\textbf {\bibinfo {volume}
			{678}},\ \bibinfo {pages} {1158} (\bibinfo {year} {2025})}\BibitemShut
	{NoStop}%
	\bibitem [{\citenamefont {Yu}\ \emph {et~al.}(2024)\citenamefont {Yu},
		\citenamefont {Ji}, \citenamefont {Lyu}, \citenamefont {Sui}, \citenamefont
		{Hao},\ and\ \citenamefont {Xu}}]{Yu2024VersatileAssemblies}%
	\BibitemOpen
	\bibfield  {author} {\bibinfo {author} {\bibfnamefont {W.}~\bibnamefont
			{Yu}}, \bibinfo {author} {\bibfnamefont {Z.}~\bibnamefont {Ji}}, \bibinfo
		{author} {\bibfnamefont {Y.}~\bibnamefont {Lyu}}, \bibinfo {author}
		{\bibfnamefont {X.}~\bibnamefont {Sui}}, \bibinfo {author} {\bibfnamefont
			{J.}~\bibnamefont {Hao}}, \ and\ \bibinfo {author} {\bibfnamefont
			{L.}~\bibnamefont {Xu}},\ }\href {\doibase 10.1039/d4mh00364k} {\bibfield
		{journal} {\bibinfo  {journal} {Materials Horizons}\ }\textbf {\bibinfo
			{volume} {11}},\ \bibinfo {pages} {3298} (\bibinfo {year}
		{2024})}\BibitemShut {NoStop}%
	\bibitem [{\citenamefont {Wu}\ \emph {et~al.}(2021)\citenamefont {Wu},
		\citenamefont {Yang}, \citenamefont {Xin}, \citenamefont {Kong},
		\citenamefont {Eggersdorfer}, \citenamefont {Ruan}, \citenamefont {Zhao},
		\citenamefont {Shan}, \citenamefont {Liu}, \citenamefont {Chen},
		\citenamefont {Weitz},\ and\ \citenamefont {Gao}}]{Wu2021AttractiveGels}%
	\BibitemOpen
	\bibfield  {author} {\bibinfo {author} {\bibfnamefont {B.}~\bibnamefont
			{Wu}}, \bibinfo {author} {\bibfnamefont {C.}~\bibnamefont {Yang}}, \bibinfo
		{author} {\bibfnamefont {Q.}~\bibnamefont {Xin}}, \bibinfo {author}
		{\bibfnamefont {L.}~\bibnamefont {Kong}}, \bibinfo {author} {\bibfnamefont
			{M.}~\bibnamefont {Eggersdorfer}}, \bibinfo {author} {\bibfnamefont
			{J.}~\bibnamefont {Ruan}}, \bibinfo {author} {\bibfnamefont {P.}~\bibnamefont
			{Zhao}}, \bibinfo {author} {\bibfnamefont {J.}~\bibnamefont {Shan}}, \bibinfo
		{author} {\bibfnamefont {K.}~\bibnamefont {Liu}}, \bibinfo {author}
		{\bibfnamefont {D.}~\bibnamefont {Chen}}, \bibinfo {author} {\bibfnamefont
			{D.~A.}\ \bibnamefont {Weitz}}, \ and\ \bibinfo {author} {\bibfnamefont
			{X.}~\bibnamefont {Gao}},\ }\href {\doibase 10.1002/adma.202102362}
	{\bibfield  {journal} {\bibinfo  {journal} {Advanced Materials}\ }\textbf
		{\bibinfo {volume} {33}},\ \bibinfo {pages} {2102362} (\bibinfo {year}
		{2021})}\BibitemShut {NoStop}%
	\bibitem [{\citenamefont {Zhang}\ \emph {et~al.}(2024)\citenamefont {Zhang},
		\citenamefont {Zheng}, \citenamefont {Guo}, \citenamefont {Sun},
		\citenamefont {Xiao}, \citenamefont {Yang}, \citenamefont {Liu},
		\citenamefont {Chen}, \citenamefont {Wu}, \citenamefont {Zhao}, \citenamefont
		{Ruan}, \citenamefont {Weitz},\ and\ \citenamefont
		{Chen}}]{Zhang2024JammedGels}%
	\BibitemOpen
	\bibfield  {author} {\bibinfo {author} {\bibfnamefont {J.}~\bibnamefont
			{Zhang}}, \bibinfo {author} {\bibfnamefont {Y.}~\bibnamefont {Zheng}},
		\bibinfo {author} {\bibfnamefont {B.}~\bibnamefont {Guo}}, \bibinfo {author}
		{\bibfnamefont {D.}~\bibnamefont {Sun}}, \bibinfo {author} {\bibfnamefont
			{Y.}~\bibnamefont {Xiao}}, \bibinfo {author} {\bibfnamefont {Z.}~\bibnamefont
			{Yang}}, \bibinfo {author} {\bibfnamefont {R.}~\bibnamefont {Liu}}, \bibinfo
		{author} {\bibfnamefont {J.}~\bibnamefont {Chen}}, \bibinfo {author}
		{\bibfnamefont {B.}~\bibnamefont {Wu}}, \bibinfo {author} {\bibfnamefont
			{P.}~\bibnamefont {Zhao}}, \bibinfo {author} {\bibfnamefont {J.}~\bibnamefont
			{Ruan}}, \bibinfo {author} {\bibfnamefont {D.~A.}\ \bibnamefont {Weitz}}, \
		and\ \bibinfo {author} {\bibfnamefont {D.}~\bibnamefont {Chen}},\ }\href
	{\doibase 10.1002/advs.202409678} {\bibfield  {journal} {\bibinfo  {journal}
			{Advanced Science}\ }\textbf {\bibinfo {volume} {11}},\ \bibinfo {pages}
		{2409678} (\bibinfo {year} {2024})}\BibitemShut {NoStop}%
	\bibitem [{\citenamefont {Wu}\ \emph {et~al.}(2024)\citenamefont {Wu},
		\citenamefont {Yuan}, \citenamefont {Fang}, \citenamefont {Wu}, \citenamefont
		{Bo}, \citenamefont {Peng},\ and\ \citenamefont
		{Wu}}]{Wu2024NaturalLubrication}%
	\BibitemOpen
	\bibfield  {author} {\bibinfo {author} {\bibfnamefont {Q.}~\bibnamefont
			{Wu}}, \bibinfo {author} {\bibfnamefont {Z.}~\bibnamefont {Yuan}}, \bibinfo
		{author} {\bibfnamefont {Y.}~\bibnamefont {Fang}}, \bibinfo {author}
		{\bibfnamefont {L.}~\bibnamefont {Wu}}, \bibinfo {author} {\bibfnamefont
			{Z.}~\bibnamefont {Bo}}, \bibinfo {author} {\bibfnamefont {C.}~\bibnamefont
			{Peng}}, \ and\ \bibinfo {author} {\bibfnamefont {B.}~\bibnamefont {Wu}},\
	}\href {\doibase 10.1016/j.colsurfb.2024.113993} {\bibfield  {journal}
		{\bibinfo  {journal} {Colloids and Surfaces B: Biointerfaces}\ }\textbf
		{\bibinfo {volume} {240}},\ \bibinfo {pages} {113993} (\bibinfo {year}
		{2024})}\BibitemShut {NoStop}%
	\bibitem [{\citenamefont {Zhang}\ \emph
		{et~al.}(2025{\natexlab{b}})\citenamefont {Zhang}, \citenamefont {Yang},
		\citenamefont {Wang}, \citenamefont {Xu}, \citenamefont {Yu},\ and\
		\citenamefont {Xu}}]{Zhang20254D-PrintableSurfactants}%
	\BibitemOpen
	\bibfield  {author} {\bibinfo {author} {\bibfnamefont {Y.}~\bibnamefont
			{Zhang}}, \bibinfo {author} {\bibfnamefont {Y.}~\bibnamefont {Yang}},
		\bibinfo {author} {\bibfnamefont {Y.}~\bibnamefont {Wang}}, \bibinfo {author}
		{\bibfnamefont {W.}~\bibnamefont {Xu}}, \bibinfo {author} {\bibfnamefont
			{W.}~\bibnamefont {Yu}}, \ and\ \bibinfo {author} {\bibfnamefont
			{L.}~\bibnamefont {Xu}},\ }\href {\doibase 10.1002/smll.202512282} {\bibfield
		{journal} {\bibinfo  {journal} {Small}\ }\textbf {\bibinfo {volume} {22}},\
		\bibinfo {pages} {e12282} (\bibinfo {year} {2025}{\natexlab{b}})}\BibitemShut
	{NoStop}%
	\bibitem [{\citenamefont {Wu}\ \emph {et~al.}(2017)\citenamefont {Wu},
		\citenamefont {Zeng}, \citenamefont {Ren}, \citenamefont {de~Vries},\ and\
		\citenamefont {van~der Heide}}]{Wu2017TheEmulsion}%
	\BibitemOpen
	\bibfield  {author} {\bibinfo {author} {\bibfnamefont {Y.}~\bibnamefont
			{Wu}}, \bibinfo {author} {\bibfnamefont {X.}~\bibnamefont {Zeng}}, \bibinfo
		{author} {\bibfnamefont {T.}~\bibnamefont {Ren}}, \bibinfo {author}
		{\bibfnamefont {E.}~\bibnamefont {de~Vries}}, \ and\ \bibinfo {author}
		{\bibfnamefont {E.}~\bibnamefont {van~der Heide}},\ }\href {\doibase
		10.1016/j.triboint.2016.10.024} {\bibfield  {journal} {\bibinfo  {journal}
			{Tribology International}\ }\textbf {\bibinfo {volume} {105}},\ \bibinfo
		{pages} {304} (\bibinfo {year} {2017})}\BibitemShut {NoStop}%
	\bibitem [{\citenamefont {Yang}\ \emph {et~al.}(2019)\citenamefont {Yang},
		\citenamefont {Zhao}, \citenamefont {Xue}, \citenamefont {Deng},
		\citenamefont {Li},\ and\ \citenamefont
		{Zeng}}]{Yang2019Branch-chainEmulsions}%
	\BibitemOpen
	\bibfield  {author} {\bibinfo {author} {\bibfnamefont {H.}~\bibnamefont
			{Yang}}, \bibinfo {author} {\bibfnamefont {L.}~\bibnamefont {Zhao}}, \bibinfo
		{author} {\bibfnamefont {S.}~\bibnamefont {Xue}}, \bibinfo {author}
		{\bibfnamefont {Z.}~\bibnamefont {Deng}}, \bibinfo {author} {\bibfnamefont
			{J.}~\bibnamefont {Li}}, \ and\ \bibinfo {author} {\bibfnamefont
			{X.}~\bibnamefont {Zeng}},\ }\href {\doibase 10.1016/j.colsurfa.2019.123703}
	{\bibfield  {journal} {\bibinfo  {journal} {Colloids and Surfaces A:
				Physicochemical and Engineering Aspects}\ }\textbf {\bibinfo {volume}
			{579}},\ \bibinfo {pages} {123703} (\bibinfo {year} {2019})}\BibitemShut
	{NoStop}%
	\bibitem [{\citenamefont {Xu}\ \emph {et~al.}(2025)\citenamefont {Xu},
		\citenamefont {Song}, \citenamefont {Fan},\ and\ \citenamefont
		{Cheng}}]{Xu20252DApplication}%
	\BibitemOpen
	\bibfield  {author} {\bibinfo {author} {\bibfnamefont {Y.~Y.}\ \bibnamefont
			{Xu}}, \bibinfo {author} {\bibfnamefont {C.}~\bibnamefont {Song}}, \bibinfo
		{author} {\bibfnamefont {L.}~\bibnamefont {Fan}}, \ and\ \bibinfo {author}
		{\bibfnamefont {Z.~L.}\ \bibnamefont {Cheng}},\ }\href {\doibase
		10.1039/d4nj05170j} {\bibfield  {journal} {\bibinfo  {journal} {New Journal
				of Chemistry}\ }\textbf {\bibinfo {volume} {49}},\ \bibinfo {pages} {4320}
		(\bibinfo {year} {2025})}\BibitemShut {NoStop}%
	\bibitem [{\citenamefont {Kumar}\ \emph
		{et~al.}(2024{\natexlab{a}})\citenamefont {Kumar}, \citenamefont {Kumar},
		\citenamefont {Joshi},\ and\ \citenamefont
		{Singh}}]{Kumar2024TribologicalNanoparticles}%
	\BibitemOpen
	\bibfield  {author} {\bibinfo {author} {\bibfnamefont {A.}~\bibnamefont
			{Kumar}}, \bibinfo {author} {\bibfnamefont {V.}~\bibnamefont {Kumar}},
		\bibinfo {author} {\bibfnamefont {Y.~M.}\ \bibnamefont {Joshi}}, \ and\
		\bibinfo {author} {\bibfnamefont {M.~K.}\ \bibnamefont {Singh}},\ }\href
	{\doibase 10.1021/acs.langmuir.3c03220} {\bibfield  {journal} {\bibinfo
			{journal} {Langmuir}\ }\textbf {\bibinfo {volume} {40}},\ \bibinfo {pages}
		{7310} (\bibinfo {year} {2024}{\natexlab{a}})}\BibitemShut {NoStop}%
	\bibitem [{\citenamefont {Kumar}\ \emph {et~al.}(2025)\citenamefont {Kumar},
		\citenamefont {Kumar}, \citenamefont {Joshi},\ and\ \citenamefont
		{Singh}}]{Kumar2025Nano-silicaLubricant}%
	\BibitemOpen
	\bibfield  {author} {\bibinfo {author} {\bibfnamefont {A.}~\bibnamefont
			{Kumar}}, \bibinfo {author} {\bibfnamefont {V.}~\bibnamefont {Kumar}},
		\bibinfo {author} {\bibfnamefont {Y.~M.}\ \bibnamefont {Joshi}}, \ and\
		\bibinfo {author} {\bibfnamefont {M.~K.}\ \bibnamefont {Singh}},\ }\href
	{\doibase 10.1016/j.jcis.2025.138486} {\bibfield  {journal} {\bibinfo
			{journal} {Journal of Colloid and Interface Science}\ }\textbf {\bibinfo
			{volume} {700}},\ \bibinfo {pages} {138486} (\bibinfo {year}
		{2025})}\BibitemShut {NoStop}%
	\bibitem [{\citenamefont {Li}\ \emph {et~al.}(2025)\citenamefont {Li},
		\citenamefont {Sui}, \citenamefont {Ayala}, \citenamefont {Marquis},
		\citenamefont {Rabl}, \citenamefont {Ertl}, \citenamefont {Bilotto},
		\citenamefont {Shang}, \citenamefont {Li}, \citenamefont {Xu}, \citenamefont
		{Righi}, \citenamefont {Eder},\ and\ \citenamefont
		{Gachot}}]{Li2025AdvancedEvaluation}%
	\BibitemOpen
	\bibfield  {author} {\bibinfo {author} {\bibfnamefont {H.}~\bibnamefont
			{Li}}, \bibinfo {author} {\bibfnamefont {X.}~\bibnamefont {Sui}}, \bibinfo
		{author} {\bibfnamefont {P.}~\bibnamefont {Ayala}}, \bibinfo {author}
		{\bibfnamefont {E.}~\bibnamefont {Marquis}}, \bibinfo {author} {\bibfnamefont
			{H.}~\bibnamefont {Rabl}}, \bibinfo {author} {\bibfnamefont {A.}~\bibnamefont
			{Ertl}}, \bibinfo {author} {\bibfnamefont {P.}~\bibnamefont {Bilotto}},
		\bibinfo {author} {\bibfnamefont {Y.}~\bibnamefont {Shang}}, \bibinfo
		{author} {\bibfnamefont {J.}~\bibnamefont {Li}}, \bibinfo {author}
		{\bibfnamefont {L.}~\bibnamefont {Xu}}, \bibinfo {author} {\bibfnamefont
			{M.~C.}\ \bibnamefont {Righi}}, \bibinfo {author} {\bibfnamefont
			{D.}~\bibnamefont {Eder}}, \ and\ \bibinfo {author} {\bibfnamefont
			{C.}~\bibnamefont {Gachot}},\ }\href {\doibase 10.1002/advs.202415268}
	{\bibfield  {journal} {\bibinfo  {journal} {Advanced Science}\ }\textbf
		{\bibinfo {volume} {12}},\ \bibinfo {pages} {2415268} (\bibinfo {year}
		{2025})}\BibitemShut {NoStop}%
	\bibitem [{\citenamefont {Cionti}\ \emph {et~al.}(2022)\citenamefont {Cionti},
		\citenamefont {Vavassori}, \citenamefont {Pargoletti}, \citenamefont
		{Meroni},\ and\ \citenamefont
		{Cappelletti}}]{Cionti2022One-stepFunctionalization}%
	\BibitemOpen
	\bibfield  {author} {\bibinfo {author} {\bibfnamefont {C.}~\bibnamefont
			{Cionti}}, \bibinfo {author} {\bibfnamefont {G.}~\bibnamefont {Vavassori}},
		\bibinfo {author} {\bibfnamefont {E.}~\bibnamefont {Pargoletti}}, \bibinfo
		{author} {\bibfnamefont {D.}~\bibnamefont {Meroni}}, \ and\ \bibinfo {author}
		{\bibfnamefont {G.}~\bibnamefont {Cappelletti}},\ }\href {\doibase
		10.1016/j.jcis.2022.07.129} {\bibfield  {journal} {\bibinfo  {journal}
			{Journal of Colloid and Interface Science}\ }\textbf {\bibinfo {volume}
			{628}},\ \bibinfo {pages} {82} (\bibinfo {year} {2022})}\BibitemShut
	{NoStop}%
	\bibitem [{\citenamefont {Silmore}\ \emph {et~al.}(2016)\citenamefont
		{Silmore}, \citenamefont {Gupta},\ and\ \citenamefont
		{Washburn}}]{Silmore2016TunablePGLNs}%
	\BibitemOpen
	\bibfield  {author} {\bibinfo {author} {\bibfnamefont {K.~S.}\ \bibnamefont
			{Silmore}}, \bibinfo {author} {\bibfnamefont {C.}~\bibnamefont {Gupta}}, \
		and\ \bibinfo {author} {\bibfnamefont {N.~R.}\ \bibnamefont {Washburn}},\
	}\href {\doibase 10.1016/j.jcis.2015.11.042} {\bibfield  {journal} {\bibinfo
			{journal} {Journal of Colloid and Interface Science}\ }\textbf {\bibinfo
			{volume} {466}},\ \bibinfo {pages} {91} (\bibinfo {year} {2016})}\BibitemShut
	{NoStop}%
	\bibitem [{\citenamefont {You}\ \emph {et~al.}(2023)\citenamefont {You},
		\citenamefont {Murray},\ and\ \citenamefont
		{Sarkar}}]{You2023TribologyMicrogels}%
	\BibitemOpen
	\bibfield  {author} {\bibinfo {author} {\bibfnamefont {K.~M.}\ \bibnamefont
			{You}}, \bibinfo {author} {\bibfnamefont {B.~S.}\ \bibnamefont {Murray}}, \
		and\ \bibinfo {author} {\bibfnamefont {A.}~\bibnamefont {Sarkar}},\ }\href
	{\doibase 10.1016/j.foodhyd.2022.108009} {\bibfield  {journal} {\bibinfo
			{journal} {Food Hydrocolloids}\ }\textbf {\bibinfo {volume} {134}},\ \bibinfo
		{pages} {108009} (\bibinfo {year} {2023})}\BibitemShut {NoStop}%
	\bibitem [{\citenamefont {Bertsch}\ \emph {et~al.}(2025)\citenamefont
		{Bertsch}, \citenamefont {Fr{\o}slev}, \citenamefont {Currie}, \citenamefont
		{Carri{\`{e}}re}, \citenamefont {M{\"{u}}llertz},\ and\ \citenamefont
		{Nielsen}}]{Bertsch2025PickeringEnhancers}%
	\BibitemOpen
	\bibfield  {author} {\bibinfo {author} {\bibfnamefont {P.}~\bibnamefont
			{Bertsch}}, \bibinfo {author} {\bibfnamefont {P.}~\bibnamefont {Fr{\o}slev}},
		\bibinfo {author} {\bibfnamefont {J.}~\bibnamefont {Currie}}, \bibinfo
		{author} {\bibfnamefont {F.}~\bibnamefont {Carri{\`{e}}re}}, \bibinfo
		{author} {\bibfnamefont {A.}~\bibnamefont {M{\"{u}}llertz}}, \ and\ \bibinfo
		{author} {\bibfnamefont {H.~M.}\ \bibnamefont {Nielsen}},\ }\href {\doibase
		10.1016/j.jcis.2025.138363} {\bibfield  {journal} {\bibinfo  {journal}
			{Journal of Colloid and Interface Science}\ }\textbf {\bibinfo {volume}
			{700}},\ \bibinfo {pages} {138363} (\bibinfo {year} {2025})}\BibitemShut
	{NoStop}%
	\bibitem [{\citenamefont {Huang}\ \emph {et~al.}(2025)\citenamefont {Huang},
		\citenamefont {Zhang}, \citenamefont {Yu}, \citenamefont {Xu}, \citenamefont
		{Deng}, \citenamefont {Zhang}, \citenamefont {Yang},\ and\ \citenamefont
		{Gao}}]{Huang2025Palladium/grapheneBenzylamine}%
	\BibitemOpen
	\bibfield  {author} {\bibinfo {author} {\bibfnamefont {H.}~\bibnamefont
			{Huang}}, \bibinfo {author} {\bibfnamefont {W.}~\bibnamefont {Zhang}},
		\bibinfo {author} {\bibfnamefont {K.}~\bibnamefont {Yu}}, \bibinfo {author}
		{\bibfnamefont {J.}~\bibnamefont {Xu}}, \bibinfo {author} {\bibfnamefont
			{Z.}~\bibnamefont {Deng}}, \bibinfo {author} {\bibfnamefont {W.}~\bibnamefont
			{Zhang}}, \bibinfo {author} {\bibfnamefont {L.}~\bibnamefont {Yang}}, \ and\
		\bibinfo {author} {\bibfnamefont {Q.}~\bibnamefont {Gao}},\ }\href {\doibase
		10.1016/j.jcis.2025.138300} {\bibfield  {journal} {\bibinfo  {journal}
			{Journal of Colloid and Interface Science}\ }\textbf {\bibinfo {volume}
			{699}},\ \bibinfo {pages} {138300} (\bibinfo {year} {2025})}\BibitemShut
	{NoStop}%
	\bibitem [{\citenamefont {Zhao}\ \emph {et~al.}(2025)\citenamefont {Zhao},
		\citenamefont {Zhou}, \citenamefont {Liu}, \citenamefont {Hao}, \citenamefont
		{Han}, \citenamefont {Pan}, \citenamefont {Yuan},\ and\ \citenamefont
		{Pan}}]{Zhao2025Dumbbell-shapedEmulsions}%
	\BibitemOpen
	\bibfield  {author} {\bibinfo {author} {\bibfnamefont {N.}~\bibnamefont
			{Zhao}}, \bibinfo {author} {\bibfnamefont {C.}~\bibnamefont {Zhou}}, \bibinfo
		{author} {\bibfnamefont {W.}~\bibnamefont {Liu}}, \bibinfo {author}
		{\bibfnamefont {F.}~\bibnamefont {Hao}}, \bibinfo {author} {\bibfnamefont
			{M.}~\bibnamefont {Han}}, \bibinfo {author} {\bibfnamefont {Z.}~\bibnamefont
			{Pan}}, \bibinfo {author} {\bibfnamefont {J.}~\bibnamefont {Yuan}}, \ and\
		\bibinfo {author} {\bibfnamefont {M.}~\bibnamefont {Pan}},\ }\href {\doibase
		10.1016/j.jcis.2025.138001} {\bibfield  {journal} {\bibinfo  {journal}
			{Journal of Colloid and Interface Science}\ }\textbf {\bibinfo {volume}
			{697}},\ \bibinfo {pages} {138001} (\bibinfo {year} {2025})}\BibitemShut
	{NoStop}%
	\bibitem [{\citenamefont {Stehl}\ \emph {et~al.}(2020)\citenamefont {Stehl},
		\citenamefont {Skale}, \citenamefont {Hohl}, \citenamefont {Lvov},
		\citenamefont {Koetz}, \citenamefont {Kraume}, \citenamefont {Drews},\ and\
		\citenamefont {Von~Klitzing}}]{Stehl2020Oil-in-WaterFilterability}%
	\BibitemOpen
	\bibfield  {author} {\bibinfo {author} {\bibfnamefont {D.}~\bibnamefont
			{Stehl}}, \bibinfo {author} {\bibfnamefont {T.}~\bibnamefont {Skale}},
		\bibinfo {author} {\bibfnamefont {L.}~\bibnamefont {Hohl}}, \bibinfo {author}
		{\bibfnamefont {Y.}~\bibnamefont {Lvov}}, \bibinfo {author} {\bibfnamefont
			{J.}~\bibnamefont {Koetz}}, \bibinfo {author} {\bibfnamefont
			{M.}~\bibnamefont {Kraume}}, \bibinfo {author} {\bibfnamefont
			{A.}~\bibnamefont {Drews}}, \ and\ \bibinfo {author} {\bibfnamefont
			{R.}~\bibnamefont {Von~Klitzing}},\ }\href {\doibase 10.1021/acsanm.0c02205}
	{\bibfield  {journal} {\bibinfo  {journal} {ACS Applied Nano Materials}\
		}\textbf {\bibinfo {volume} {3}},\ \bibinfo {pages} {11743} (\bibinfo {year}
		{2020})}\BibitemShut {NoStop}%
	\bibitem [{\citenamefont {Merad}\ \emph {et~al.}(2021)\citenamefont {Merad},
		\citenamefont {Bekkour}, \citenamefont {Fran{\c{c}}ois}, \citenamefont
		{Gareche},\ and\ \citenamefont {Lawniczak}}]{Merad2021RheologicalClay}%
	\BibitemOpen
	\bibfield  {author} {\bibinfo {author} {\bibfnamefont {B.}~\bibnamefont
			{Merad}}, \bibinfo {author} {\bibfnamefont {K.}~\bibnamefont {Bekkour}},
		\bibinfo {author} {\bibfnamefont {P.}~\bibnamefont {Fran{\c{c}}ois}},
		\bibinfo {author} {\bibfnamefont {M.}~\bibnamefont {Gareche}}, \ and\
		\bibinfo {author} {\bibfnamefont {F.}~\bibnamefont {Lawniczak}},\ }\href
	{\doibase 10.1016/j.petrol.2021.108780} {\bibfield  {journal} {\bibinfo
			{journal} {Journal of Petroleum Science and Engineering}\ }\textbf {\bibinfo
			{volume} {205}},\ \bibinfo {pages} {108780} (\bibinfo {year}
		{2021})}\BibitemShut {NoStop}%
	\bibitem [{\citenamefont {Assun{\c{c}}{\~{a}}o~Dorigon}\ \emph
		{et~al.}(2025)\citenamefont {Assun{\c{c}}{\~{a}}o~Dorigon}, \citenamefont
		{Ara{\'{u}}jo~Pessoa}, \citenamefont {Queiroz~Neto}, \citenamefont
		{Andrade~Silva}, \citenamefont {Soares~Filho},\ and\ \citenamefont
		{Silva~Curbelo}}]{AssuncaoDorigon2025DevelopmentNanoclay}%
	\BibitemOpen
	\bibfield  {author} {\bibinfo {author} {\bibfnamefont {I.~C.}\ \bibnamefont
			{Assun{\c{c}}{\~{a}}o~Dorigon}}, \bibinfo {author} {\bibfnamefont {M.~E.}\
			\bibnamefont {Ara{\'{u}}jo~Pessoa}}, \bibinfo {author} {\bibfnamefont
			{J.~C.}\ \bibnamefont {Queiroz~Neto}}, \bibinfo {author} {\bibfnamefont
			{D.~V.}\ \bibnamefont {Andrade~Silva}}, \bibinfo {author} {\bibfnamefont
			{J.~E.}\ \bibnamefont {Soares~Filho}}, \ and\ \bibinfo {author}
		{\bibfnamefont {F.~D.}\ \bibnamefont {Silva~Curbelo}},\ }\href {\doibase
		10.1016/j.clay.2025.107780} {\bibfield  {journal} {\bibinfo  {journal}
			{Applied Clay Science}\ }\textbf {\bibinfo {volume} {270}},\ \bibinfo {pages}
		{107780} (\bibinfo {year} {2025})}\BibitemShut {NoStop}%
	\bibitem [{\citenamefont {Kang}\ \emph {et~al.}(2024)\citenamefont {Kang},
		\citenamefont {Zan}, \citenamefont {Cong}, \citenamefont {Wang},
		\citenamefont {Luo},\ and\ \citenamefont {Li}}]{Kang2024AActivity}%
	\BibitemOpen
	\bibfield  {author} {\bibinfo {author} {\bibfnamefont {Y.}~\bibnamefont
			{Kang}}, \bibinfo {author} {\bibfnamefont {Y.}~\bibnamefont {Zan}}, \bibinfo
		{author} {\bibfnamefont {Y.}~\bibnamefont {Cong}}, \bibinfo {author}
		{\bibfnamefont {X.}~\bibnamefont {Wang}}, \bibinfo {author} {\bibfnamefont
			{Y.}~\bibnamefont {Luo}}, \ and\ \bibinfo {author} {\bibfnamefont
			{L.}~\bibnamefont {Li}},\ }\href {\doibase 10.1016/j.colsurfa.2024.133337}
	{\bibfield  {journal} {\bibinfo  {journal} {Colloids and Surfaces A:
				Physicochemical and Engineering Aspects}\ }\textbf {\bibinfo {volume}
			{686}},\ \bibinfo {pages} {133337} (\bibinfo {year} {2024})}\BibitemShut
	{NoStop}%
	\bibitem [{\citenamefont {Lu}\ \emph {et~al.}(2021{\natexlab{b}})\citenamefont
		{Lu}, \citenamefont {Gou}, \citenamefont {Rao},\ and\ \citenamefont
		{Zhao}}]{Lu2021RecentApplications}%
	\BibitemOpen
	\bibfield  {author} {\bibinfo {author} {\bibfnamefont {T.}~\bibnamefont
			{Lu}}, \bibinfo {author} {\bibfnamefont {H.}~\bibnamefont {Gou}}, \bibinfo
		{author} {\bibfnamefont {H.}~\bibnamefont {Rao}}, \ and\ \bibinfo {author}
		{\bibfnamefont {G.}~\bibnamefont {Zhao}},\ }\href {\doibase
		10.1016/j.jece.2021.105941} {\bibfield  {journal} {\bibinfo  {journal}
			{Journal of Environmental Chemical Engineering}\ }\textbf {\bibinfo {volume}
			{9}},\ \bibinfo {pages} {105941} (\bibinfo {year}
		{2021}{\natexlab{b}})}\BibitemShut {NoStop}%
	\bibitem [{\citenamefont {Lisuzzo}\ \emph {et~al.}(2022)\citenamefont
		{Lisuzzo}, \citenamefont {Cavallaro}, \citenamefont {Milioto},\ and\
		\citenamefont {Lazzara}}]{Lisuzzo2022PickeringApplications}%
	\BibitemOpen
	\bibfield  {author} {\bibinfo {author} {\bibfnamefont {L.}~\bibnamefont
			{Lisuzzo}}, \bibinfo {author} {\bibfnamefont {G.}~\bibnamefont {Cavallaro}},
		\bibinfo {author} {\bibfnamefont {S.}~\bibnamefont {Milioto}}, \ and\
		\bibinfo {author} {\bibfnamefont {G.}~\bibnamefont {Lazzara}},\ }\href
	{\doibase 10.1002/admi.202102346} {\bibfield  {journal} {\bibinfo  {journal}
			{Advanced Materials Interfaces}\ }\textbf {\bibinfo {volume} {9}},\ \bibinfo
		{pages} {2102346} (\bibinfo {year} {2022})}\BibitemShut {NoStop}%
	\bibitem [{\citenamefont {Liu}\ \emph {et~al.}(2021)\citenamefont {Liu},
		\citenamefont {Pu}, \citenamefont {Tao}, \citenamefont {Chen}, \citenamefont
		{Guo}, \citenamefont {Luo},\ and\ \citenamefont
		{Ren}}]{Liu2021PickeringConditions}%
	\BibitemOpen
	\bibfield  {author} {\bibinfo {author} {\bibfnamefont {L.}~\bibnamefont
			{Liu}}, \bibinfo {author} {\bibfnamefont {X.}~\bibnamefont {Pu}}, \bibinfo
		{author} {\bibfnamefont {H.}~\bibnamefont {Tao}}, \bibinfo {author}
		{\bibfnamefont {K.}~\bibnamefont {Chen}}, \bibinfo {author} {\bibfnamefont
			{W.}~\bibnamefont {Guo}}, \bibinfo {author} {\bibfnamefont {D.}~\bibnamefont
			{Luo}}, \ and\ \bibinfo {author} {\bibfnamefont {Z.}~\bibnamefont {Ren}},\
	}\href {\doibase 10.1016/j.colsurfa.2020.125694} {\bibfield  {journal}
		{\bibinfo  {journal} {Colloids and Surfaces A: Physicochemical and
				Engineering Aspects}\ }\textbf {\bibinfo {volume} {610}},\ \bibinfo {pages}
		{125694} (\bibinfo {year} {2021})}\BibitemShut {NoStop}%
	\bibitem [{\citenamefont {Luo}\ \emph {et~al.}(2025)\citenamefont {Luo},
		\citenamefont {Yu}, \citenamefont {Wang},\ and\ \citenamefont
		{Yu}}]{Luo2025RheologicalFluids}%
	\BibitemOpen
	\bibfield  {author} {\bibinfo {author} {\bibfnamefont {A.}~\bibnamefont
			{Luo}}, \bibinfo {author} {\bibfnamefont {L.}~\bibnamefont {Yu}}, \bibinfo
		{author} {\bibfnamefont {X.}~\bibnamefont {Wang}}, \ and\ \bibinfo {author}
		{\bibfnamefont {P.}~\bibnamefont {Yu}},\ }\href {\doibase
		10.1016/j.geoen.2024.213418} {\bibfield  {journal} {\bibinfo  {journal}
			{Geoenergy Science and Engineering}\ }\textbf {\bibinfo {volume} {244}},\
		\bibinfo {pages} {213418} (\bibinfo {year} {2025})}\BibitemShut {NoStop}%
	\bibitem [{\citenamefont {Zheng}\ \emph {et~al.}(2020)\citenamefont {Zheng},
		\citenamefont {Zheng}, \citenamefont {Carr}, \citenamefont {Yu},
		\citenamefont {McClements},\ and\ \citenamefont
		{Bhatia}}]{Zheng2020EmulsionsApplications}%
	\BibitemOpen
	\bibfield  {author} {\bibinfo {author} {\bibfnamefont {B.}~\bibnamefont
			{Zheng}}, \bibinfo {author} {\bibfnamefont {B.}~\bibnamefont {Zheng}},
		\bibinfo {author} {\bibfnamefont {A.~J.}\ \bibnamefont {Carr}}, \bibinfo
		{author} {\bibfnamefont {X.}~\bibnamefont {Yu}}, \bibinfo {author}
		{\bibfnamefont {D.~J.}\ \bibnamefont {McClements}}, \ and\ \bibinfo {author}
		{\bibfnamefont {S.~R.}\ \bibnamefont {Bhatia}},\ }\href {\doibase
		10.1016/j.ica.2020.119566} {\bibfield  {journal} {\bibinfo  {journal}
			{Inorganica Chimica Acta}\ }\textbf {\bibinfo {volume} {508}},\ \bibinfo
		{pages} {119566} (\bibinfo {year} {2020})}\BibitemShut {NoStop}%
	\bibitem [{\citenamefont {Yu}\ \emph {et~al.}(2021)\citenamefont {Yu},
		\citenamefont {Li}, \citenamefont {Stubbs},\ and\ \citenamefont
		{Lau}}]{Yu2021CharacterizationReservoirs}%
	\BibitemOpen
	\bibfield  {author} {\bibinfo {author} {\bibfnamefont {L.}~\bibnamefont
			{Yu}}, \bibinfo {author} {\bibfnamefont {S.}~\bibnamefont {Li}}, \bibinfo
		{author} {\bibfnamefont {L.~P.}\ \bibnamefont {Stubbs}}, \ and\ \bibinfo
		{author} {\bibfnamefont {H.~C.}\ \bibnamefont {Lau}},\ }\href {\doibase
		10.1016/j.clay.2021.106246} {\bibfield  {journal} {\bibinfo  {journal}
			{Applied Clay Science}\ }\textbf {\bibinfo {volume} {213}},\ \bibinfo {pages}
		{106246} (\bibinfo {year} {2021})}\BibitemShut {NoStop}%
	\bibitem [{\citenamefont {Taleb}\ \emph {et~al.}(2024)\citenamefont {Taleb},
		\citenamefont {Abbou}, \citenamefont {Raho},\ and\ \citenamefont
		{Saidi-Besbes}}]{Taleb2024ComparativeEmulsion}%
	\BibitemOpen
	\bibfield  {author} {\bibinfo {author} {\bibfnamefont {K.}~\bibnamefont
			{Taleb}}, \bibinfo {author} {\bibfnamefont {I.}~\bibnamefont {Abbou}},
		\bibinfo {author} {\bibfnamefont {R.}~\bibnamefont {Raho}}, \ and\ \bibinfo
		{author} {\bibfnamefont {S.}~\bibnamefont {Saidi-Besbes}},\ }\href {\doibase
		10.1080/01932691.2024.2428341} {\bibfield  {journal} {\bibinfo  {journal}
			{Journal of Dispersion Science and Technology}\ }\textbf {\bibinfo {volume}
			{47}},\ \bibinfo {pages} {925} (\bibinfo {year} {2024})}\BibitemShut
	{NoStop}%
	\bibitem [{\citenamefont {Schubel}\ \emph {et~al.}(2006)\citenamefont
		{Schubel}, \citenamefont {Johnson}, \citenamefont {Warrior},\ and\
		\citenamefont {Rudd}}]{Schubel2006CharacterisationReinforcement}%
	\BibitemOpen
	\bibfield  {author} {\bibinfo {author} {\bibfnamefont {P.~J.}\ \bibnamefont
			{Schubel}}, \bibinfo {author} {\bibfnamefont {M.~S.}\ \bibnamefont
			{Johnson}}, \bibinfo {author} {\bibfnamefont {N.~A.}\ \bibnamefont
			{Warrior}}, \ and\ \bibinfo {author} {\bibfnamefont {C.~D.}\ \bibnamefont
			{Rudd}},\ }\href {\doibase 10.1016/j.compositesa.2005.09.014} {\bibfield
		{journal} {\bibinfo  {journal} {Composites Part A: Applied Science and
				Manufacturing}\ }\textbf {\bibinfo {volume} {37}},\ \bibinfo {pages} {1757}
		(\bibinfo {year} {2006})}\BibitemShut {NoStop}%
	\bibitem [{\citenamefont {Sarathi}\ \emph {et~al.}(2007)\citenamefont
		{Sarathi}, \citenamefont {Sahu},\ and\ \citenamefont
		{Rajeshkumar}}]{Sarathi2007UnderstandingNanocomposites}%
	\BibitemOpen
	\bibfield  {author} {\bibinfo {author} {\bibfnamefont {R.}~\bibnamefont
			{Sarathi}}, \bibinfo {author} {\bibfnamefont {R.~K.}\ \bibnamefont {Sahu}}, \
		and\ \bibinfo {author} {\bibfnamefont {P.}~\bibnamefont {Rajeshkumar}},\
	}\href {\doibase 10.1016/j.msea.2006.09.077} {\bibfield  {journal} {\bibinfo
			{journal} {Materials Science and Engineering: A}\ }\textbf {\bibinfo {volume}
			{445-446}},\ \bibinfo {pages} {567} (\bibinfo {year} {2007})}\BibitemShut
	{NoStop}%
	\bibitem [{\citenamefont {Rana}\ \emph {et~al.}(2021)\citenamefont {Rana},
		\citenamefont {Vamshi}, \citenamefont {Naresh}, \citenamefont {Velmurugan},\
		and\ \citenamefont {Sarathi}}]{Rana2021EffectComposites}%
	\BibitemOpen
	\bibfield  {author} {\bibinfo {author} {\bibfnamefont {A.~S.}\ \bibnamefont
			{Rana}}, \bibinfo {author} {\bibfnamefont {M.~K.}\ \bibnamefont {Vamshi}},
		\bibinfo {author} {\bibfnamefont {K.}~\bibnamefont {Naresh}}, \bibinfo
		{author} {\bibfnamefont {R.}~\bibnamefont {Velmurugan}}, \ and\ \bibinfo
		{author} {\bibfnamefont {R.}~\bibnamefont {Sarathi}},\ }\href {\doibase
		10.1080/2374068X.2020.1754720} {\bibfield  {journal} {\bibinfo  {journal}
			{Advances in Materials and Processing Technologies}\ }\textbf {\bibinfo
			{volume} {7}},\ \bibinfo {pages} {109} (\bibinfo {year} {2021})}\BibitemShut
	{NoStop}%
	\bibitem [{\citenamefont {Ho}\ \emph {et~al.}(2006)\citenamefont {Ho},
		\citenamefont {Lam}, \citenamefont {Lau}, \citenamefont {Ng},\ and\
		\citenamefont {Hui}}]{Ho2006MechanicalNanoclays}%
	\BibitemOpen
	\bibfield  {author} {\bibinfo {author} {\bibfnamefont {M.~W.}\ \bibnamefont
			{Ho}}, \bibinfo {author} {\bibfnamefont {C.~K.}\ \bibnamefont {Lam}},
		\bibinfo {author} {\bibfnamefont {K.~t.}\ \bibnamefont {Lau}}, \bibinfo
		{author} {\bibfnamefont {D.~H.}\ \bibnamefont {Ng}}, \ and\ \bibinfo {author}
		{\bibfnamefont {D.}~\bibnamefont {Hui}},\ }\href {\doibase
		10.1016/j.compstruct.2006.04.051} {\bibfield  {journal} {\bibinfo  {journal}
			{Composite Structures}\ }\textbf {\bibinfo {volume} {75}},\ \bibinfo {pages}
		{415} (\bibinfo {year} {2006})}\BibitemShut {NoStop}%
	\bibitem [{\citenamefont {Liu}\ \emph {et~al.}(2026)\citenamefont {Liu},
		\citenamefont {Huang}, \citenamefont {Lai},\ and\ \citenamefont
		{Chung}}]{Liu2026Garamite-reinforcedDistillation}%
	\BibitemOpen
	\bibfield  {author} {\bibinfo {author} {\bibfnamefont {H.~Y.}\ \bibnamefont
			{Liu}}, \bibinfo {author} {\bibfnamefont {Y.~H.}\ \bibnamefont {Huang}},
		\bibinfo {author} {\bibfnamefont {J.~Y.}\ \bibnamefont {Lai}}, \ and\
		\bibinfo {author} {\bibfnamefont {T.~S.}\ \bibnamefont {Chung}},\ }\href
	{\doibase 10.1016/j.memsci.2025.124778} {\bibfield  {journal} {\bibinfo
			{journal} {Journal of Membrane Science}\ }\textbf {\bibinfo {volume} {737}},\
		\bibinfo {pages} {124778} (\bibinfo {year} {2026})}\BibitemShut {NoStop}%
	\bibitem [{\citenamefont {{BYK}}(2019)}]{BYK2019RheologyAdditives}%
	\BibitemOpen
	\bibfield  {author} {\bibinfo {author} {\bibnamefont {{BYK}}},\ }\href@noop
	{} {\enquote {\bibinfo {title} {{Rheology additives}},}\ } (\bibinfo {year}
	{2019})\BibitemShut {NoStop}%
	\bibitem [{\citenamefont {Panda}\ \emph {et~al.}(2025)\citenamefont {Panda},
		\citenamefont {Maity}, \citenamefont {Dutta},\ and\ \citenamefont
		{Das}}]{Panda2025AnisotropicActuation}%
	\BibitemOpen
	\bibfield  {author} {\bibinfo {author} {\bibfnamefont {P.}~\bibnamefont
			{Panda}}, \bibinfo {author} {\bibfnamefont {P.}~\bibnamefont {Maity}},
		\bibinfo {author} {\bibfnamefont {A.}~\bibnamefont {Dutta}}, \ and\ \bibinfo
		{author} {\bibfnamefont {R.~K.}\ \bibnamefont {Das}},\ }\href {\doibase
		10.1021/acs.langmuir.5c01032} {\bibfield  {journal} {\bibinfo  {journal}
			{Langmuir}\ }\textbf {\bibinfo {volume} {41}},\ \bibinfo {pages} {13301}
		(\bibinfo {year} {2025})}\BibitemShut {NoStop}%
	\bibitem [{\citenamefont {Teixema}(1988)}]{Teixema1988Small-AngleSystems}%
	\BibitemOpen
	\bibfield  {author} {\bibinfo {author} {\bibfnamefont {J.}~\bibnamefont
			{Teixema}},\ }\href@noop {} {\bibfield  {journal} {\bibinfo  {journal} {J.
				Appl. Cryst}\ }\textbf {\bibinfo {volume} {21}},\ \bibinfo {pages} {781}
		(\bibinfo {year} {1988})}\BibitemShut {NoStop}%
	\bibitem [{\citenamefont {Kong}\ \emph {et~al.}(2025)\citenamefont {Kong},
		\citenamefont {Ma}, \citenamefont {Zhen}, \citenamefont {Liu}, \citenamefont
		{Sun},\ and\ \citenamefont {Yang}}]{Kong2025ElucidatingMayonnaise}%
	\BibitemOpen
	\bibfield  {author} {\bibinfo {author} {\bibfnamefont {S.}~\bibnamefont
			{Kong}}, \bibinfo {author} {\bibfnamefont {X.}~\bibnamefont {Ma}}, \bibinfo
		{author} {\bibfnamefont {S.}~\bibnamefont {Zhen}}, \bibinfo {author}
		{\bibfnamefont {Y.}~\bibnamefont {Liu}}, \bibinfo {author} {\bibfnamefont
			{F.}~\bibnamefont {Sun}}, \ and\ \bibinfo {author} {\bibfnamefont
			{N.}~\bibnamefont {Yang}},\ }\href {\doibase 10.1016/j.ijbiomac.2025.139650}
	{\bibfield  {journal} {\bibinfo  {journal} {International Journal of
				Biological Macromolecules}\ }\textbf {\bibinfo {volume} {296}},\ \bibinfo
		{pages} {139650} (\bibinfo {year} {2025})}\BibitemShut {NoStop}%
	\bibitem [{\citenamefont {Thompson}\ \emph {et~al.}(2001)\citenamefont
		{Thompson}, \citenamefont {Capehart},\ and\ \citenamefont
		{Fuller}}]{Thompson2001ConfinementMelting}%
	\BibitemOpen
	\bibfield  {author} {\bibinfo {author} {\bibfnamefont {E.~L.}\ \bibnamefont
			{Thompson}}, \bibinfo {author} {\bibfnamefont {T.~W.}\ \bibnamefont
			{Capehart}}, \ and\ \bibinfo {author} {\bibfnamefont {T.~J.}\ \bibnamefont
			{Fuller}},\ }\href
	{https://iopscience.iop.org/article/10.1088/0953-8984/13/11/201/pdf}
	{\bibfield  {journal} {\bibinfo  {journal} {J. Phys.: Condens. Matter}\
		}\textbf {\bibinfo {volume} {13}},\ \bibinfo {pages} {95} (\bibinfo {year}
		{2001})}\BibitemShut {NoStop}%
	\bibitem [{\citenamefont {Morishige}\ and\ \citenamefont
		{Kawano}(1999)}]{Morishige1999FreezingBehavior}%
	\BibitemOpen
	\bibfield  {author} {\bibinfo {author} {\bibfnamefont {K.}~\bibnamefont
			{Morishige}}\ and\ \bibinfo {author} {\bibfnamefont {K.}~\bibnamefont
			{Kawano}},\ }\href {\doibase 10.1063/1.478372} {\bibfield  {journal}
		{\bibinfo  {journal} {Journal of Chemical Physics}\ }\textbf {\bibinfo
			{volume} {110}},\ \bibinfo {pages} {4867} (\bibinfo {year}
		{1999})}\BibitemShut {NoStop}%
	\bibitem [{\citenamefont {Joshi}(2014)}]{Joshi2014DynamicsGels}%
	\BibitemOpen
	\bibfield  {author} {\bibinfo {author} {\bibfnamefont {Y.~M.}\ \bibnamefont
			{Joshi}},\ }\href {\doibase 10.1146/annurev-chembioeng-060713-040230}
	{\bibfield  {journal} {\bibinfo  {journal} {Annual Review of Chemical and
				Biomolecular Engineering}\ }\textbf {\bibinfo {volume} {5}},\ \bibinfo
		{pages} {181} (\bibinfo {year} {2014})}\BibitemShut {NoStop}%
	\bibitem [{\citenamefont {Joshi}(2015)}]{Joshi2015AMaterials}%
	\BibitemOpen
	\bibfield  {author} {\bibinfo {author} {\bibfnamefont {Y.~M.}\ \bibnamefont
			{Joshi}},\ }\href {\doibase 10.1039/c5sm00217f} {\bibfield  {journal}
		{\bibinfo  {journal} {Soft Matter}\ }\textbf {\bibinfo {volume} {11}},\
		\bibinfo {pages} {3198} (\bibinfo {year} {2015})}\BibitemShut {NoStop}%
	\bibitem [{\citenamefont {Munro}\ \emph {et~al.}(2022)\citenamefont {Munro},
		\citenamefont {Hall},\ and\ \citenamefont
		{Whitby}}]{Munro2022YieldingFractions}%
	\BibitemOpen
	\bibfield  {author} {\bibinfo {author} {\bibfnamefont {B.~C.}\ \bibnamefont
			{Munro}}, \bibinfo {author} {\bibfnamefont {S.~B.}\ \bibnamefont {Hall}}, \
		and\ \bibinfo {author} {\bibfnamefont {C.~P.}\ \bibnamefont {Whitby}},\
	}\href {\doibase 10.1016/j.colsurfa.2021.128237} {\bibfield  {journal}
		{\bibinfo  {journal} {Colloids and Surfaces A: Physicochemical and
				Engineering Aspects}\ }\textbf {\bibinfo {volume} {637}},\ \bibinfo {pages}
		{128237} (\bibinfo {year} {2022})}\BibitemShut {NoStop}%
	\bibitem [{\citenamefont {Kaushal}\ and\ \citenamefont
		{Joshi}(2014)}]{Kaushal2014LinearMaterials}%
	\BibitemOpen
	\bibfield  {author} {\bibinfo {author} {\bibfnamefont {M.}~\bibnamefont
			{Kaushal}}\ and\ \bibinfo {author} {\bibfnamefont {Y.~M.}\ \bibnamefont
			{Joshi}},\ }\href {\doibase 10.1039/c3sm52978a} {\bibfield  {journal}
		{\bibinfo  {journal} {Soft Matter}\ }\textbf {\bibinfo {volume} {10}},\
		\bibinfo {pages} {1891} (\bibinfo {year} {2014})}\BibitemShut {NoStop}%
	\bibitem [{\citenamefont {Keane}\ \emph {et~al.}(2025)\citenamefont {Keane},
		\citenamefont {Nikoumanesh}, \citenamefont {Kamani}, \citenamefont {Rogers},\
		and\ \citenamefont {Poling-Skutvik}}]{Keane2025UniversalMaterials}%
	\BibitemOpen
	\bibfield  {author} {\bibinfo {author} {\bibfnamefont {D.~P.}\ \bibnamefont
			{Keane}}, \bibinfo {author} {\bibfnamefont {E.}~\bibnamefont {Nikoumanesh}},
		\bibinfo {author} {\bibfnamefont {K.~M.}\ \bibnamefont {Kamani}}, \bibinfo
		{author} {\bibfnamefont {S.~A.}\ \bibnamefont {Rogers}}, \ and\ \bibinfo
		{author} {\bibfnamefont {R.}~\bibnamefont {Poling-Skutvik}},\ }\href
	{\doibase 10.1103/PhysRevLett.134.208202} {\bibfield  {journal} {\bibinfo
			{journal} {Physical Review Letters}\ }\textbf {\bibinfo {volume} {134}},\
		\bibinfo {pages} {208202} (\bibinfo {year} {2025})}\BibitemShut {NoStop}%
	\bibitem [{\citenamefont {Joshi}\ and\ \citenamefont
		{Petekidis}(2018)}]{Joshi2018YieldAgeing}%
	\BibitemOpen
	\bibfield  {author} {\bibinfo {author} {\bibfnamefont {Y.~M.}\ \bibnamefont
			{Joshi}}\ and\ \bibinfo {author} {\bibfnamefont {G.}~\bibnamefont
			{Petekidis}},\ }\href {\doibase 10.1007/s00397-018-1096-6} {\bibfield
		{journal} {\bibinfo  {journal} {Rheologica Acta}\ }\textbf {\bibinfo {volume}
			{57}},\ \bibinfo {pages} {521} (\bibinfo {year} {2018})}\BibitemShut
	{NoStop}%
	\bibitem [{\citenamefont {Ciccone}\ \emph {et~al.}(2022)\citenamefont
		{Ciccone}, \citenamefont {Skopalik}, \citenamefont {Smart}, \citenamefont
		{Gezgin}, \citenamefont {Ridland}, \citenamefont {Paul}, \citenamefont
		{Noriega~Escobar},\ and\ \citenamefont
		{Tassieri}}]{Ciccone2022AFormulations}%
	\BibitemOpen
	\bibfield  {author} {\bibinfo {author} {\bibfnamefont {G.}~\bibnamefont
			{Ciccone}}, \bibinfo {author} {\bibfnamefont {S.}~\bibnamefont {Skopalik}},
		\bibinfo {author} {\bibfnamefont {C.}~\bibnamefont {Smart}}, \bibinfo
		{author} {\bibfnamefont {S.}~\bibnamefont {Gezgin}}, \bibinfo {author}
		{\bibfnamefont {D.}~\bibnamefont {Ridland}}, \bibinfo {author} {\bibfnamefont
			{M.~C.}\ \bibnamefont {Paul}}, \bibinfo {author} {\bibfnamefont {M.~D.~P.}\
			\bibnamefont {Noriega~Escobar}}, \ and\ \bibinfo {author} {\bibfnamefont
			{M.}~\bibnamefont {Tassieri}},\ }\href {\doibase 10.1063/5.0099145}
	{\bibfield  {journal} {\bibinfo  {journal} {Physics of Fluids}\ }\textbf
		{\bibinfo {volume} {34}},\ \bibinfo {pages} {097109} (\bibinfo {year}
		{2022})}\BibitemShut {NoStop}%
	\bibitem [{\citenamefont {Majumder}\ \emph {et~al.}(2026)\citenamefont
		{Majumder}, \citenamefont {Dunlop}, \citenamefont {Acharya},\ and\
		\citenamefont {Ghosh}}]{Majumder2026RheologyMethod}%
	\BibitemOpen
	\bibfield  {author} {\bibinfo {author} {\bibfnamefont {S.}~\bibnamefont
			{Majumder}}, \bibinfo {author} {\bibfnamefont {M.~J.}\ \bibnamefont
			{Dunlop}}, \bibinfo {author} {\bibfnamefont {B.}~\bibnamefont {Acharya}}, \
		and\ \bibinfo {author} {\bibfnamefont {S.}~\bibnamefont {Ghosh}},\ }\href
	{\doibase 10.3390/foods15030509} {\bibfield  {journal} {\bibinfo  {journal}
			{Foods}\ }\textbf {\bibinfo {volume} {15}},\ \bibinfo {pages} {509} (\bibinfo
		{year} {2026})}\BibitemShut {NoStop}%
	\bibitem [{\citenamefont {Agrawal}\ and\ \citenamefont
		{Garcia-Tunon}(2024)}]{Agrawal2024InterplayRheology}%
	\BibitemOpen
	\bibfield  {author} {\bibinfo {author} {\bibfnamefont {R.}~\bibnamefont
			{Agrawal}}\ and\ \bibinfo {author} {\bibfnamefont {E.}~\bibnamefont
			{Garcia-Tunon}},\ }\href {\doibase 10.1039/d4sm00758a} {\bibfield  {journal}
		{\bibinfo  {journal} {Soft Matter}\ }\textbf {\bibinfo {volume} {20}},\
		\bibinfo {pages} {7429} (\bibinfo {year} {2024})}\BibitemShut {NoStop}%
	\bibitem [{\citenamefont {Bhattacharyya}\ \emph {et~al.}(2023)\citenamefont
		{Bhattacharyya}, \citenamefont {Jacob}, \citenamefont {Petekidis},\ and\
		\citenamefont {Joshi}}]{Bhattacharyya2023OnMaterials}%
	\BibitemOpen
	\bibfield  {author} {\bibinfo {author} {\bibfnamefont {T.}~\bibnamefont
			{Bhattacharyya}}, \bibinfo {author} {\bibfnamefont {A.~R.}\ \bibnamefont
			{Jacob}}, \bibinfo {author} {\bibfnamefont {G.}~\bibnamefont {Petekidis}}, \
		and\ \bibinfo {author} {\bibfnamefont {Y.~M.}\ \bibnamefont {Joshi}},\ }\href
	{\doibase 10.1122/8.0000558} {\bibfield  {journal} {\bibinfo  {journal}
			{Journal of Rheology}\ }\textbf {\bibinfo {volume} {67}},\ \bibinfo {pages}
		{461} (\bibinfo {year} {2023})}\BibitemShut {NoStop}%
	\bibitem [{\citenamefont {Joshi}(2025)}]{Joshi2025LinearMaterials}%
	\BibitemOpen
	\bibfield  {author} {\bibinfo {author} {\bibfnamefont {Y.~M.}\ \bibnamefont
			{Joshi}},\ }\href {\doibase 10.1016/j.cocis.2025.101896} {\bibfield
		{journal} {\bibinfo  {journal} {Current Opinion in Colloid and Interface
				Science}\ }\textbf {\bibinfo {volume} {76}},\ \bibinfo {pages} {101896}
		(\bibinfo {year} {2025})}\BibitemShut {NoStop}%
	\bibitem [{\citenamefont {Qin}\ \emph {et~al.}(2025)\citenamefont {Qin},
		\citenamefont {Yang}, \citenamefont {Lu}, \citenamefont {Li}, \citenamefont
		{Ma}, \citenamefont {Ma},\ and\ \citenamefont
		{Zhou}}]{Qin2025TribologyMaterials}%
	\BibitemOpen
	\bibfield  {author} {\bibinfo {author} {\bibfnamefont {C.}~\bibnamefont
			{Qin}}, \bibinfo {author} {\bibfnamefont {H.}~\bibnamefont {Yang}}, \bibinfo
		{author} {\bibfnamefont {Y.}~\bibnamefont {Lu}}, \bibinfo {author}
		{\bibfnamefont {B.}~\bibnamefont {Li}}, \bibinfo {author} {\bibfnamefont
			{S.}~\bibnamefont {Ma}}, \bibinfo {author} {\bibfnamefont {Y.}~\bibnamefont
			{Ma}}, \ and\ \bibinfo {author} {\bibfnamefont {F.}~\bibnamefont {Zhou}},\
	}\href {\doibase 10.1002/adma.202420626} {\bibfield  {journal} {\bibinfo
			{journal} {Advanced Materials}\ }\textbf {\bibinfo {volume} {37}},\ \bibinfo
		{pages} {2420626} (\bibinfo {year} {2025})}\BibitemShut {NoStop}%
	\bibitem [{\citenamefont {Hamrock}\ and\ \citenamefont
		{Dowson}(1977)}]{Hamrock1977IsothermalResult}%
	\BibitemOpen
	\bibfield  {author} {\bibinfo {author} {\bibfnamefont {B.~J.}\ \bibnamefont
			{Hamrock}}\ and\ \bibinfo {author} {\bibfnamefont {D.}~\bibnamefont
			{Dowson}},\ }\href {\doibase 10.1115/1.3453074} {\bibfield  {journal}
		{\bibinfo  {journal} {Journal of Tribology}\ }\textbf {\bibinfo {volume}
			{99}},\ \bibinfo {pages} {264} (\bibinfo {year} {1977})}\BibitemShut
	{NoStop}%
	\bibitem [{\citenamefont {Wang}\ \emph {et~al.}(2025)\citenamefont {Wang},
		\citenamefont {Murray}, \citenamefont {Bryant},\ and\ \citenamefont
		{Sarkar}}]{Wang2025PickeringPerformance}%
	\BibitemOpen
	\bibfield  {author} {\bibinfo {author} {\bibfnamefont {C.}~\bibnamefont
			{Wang}}, \bibinfo {author} {\bibfnamefont {B.~S.}\ \bibnamefont {Murray}},
		\bibinfo {author} {\bibfnamefont {M.}~\bibnamefont {Bryant}}, \ and\ \bibinfo
		{author} {\bibfnamefont {A.}~\bibnamefont {Sarkar}},\ }\href {\doibase
		10.1016/j.cocis.2025.101940} {\bibfield  {journal} {\bibinfo  {journal}
			{Current Opinion in Colloid and Interface Science}\ }\textbf {\bibinfo
			{volume} {79}},\ \bibinfo {pages} {101940} (\bibinfo {year}
		{2025})}\BibitemShut {NoStop}%
	\bibitem [{\citenamefont {Miller}(2011)}]{Miller2011TWOCOMPOSITIONS}%
	\BibitemOpen
	\bibfield  {author} {\bibinfo {author} {\bibfnamefont {M.}~\bibnamefont
			{Miller}},\ }\href@noop {} {\enquote {\bibinfo {title} {{Two component
					curable compositions}},}\ } (\bibinfo {year} {2011})\BibitemShut {NoStop}%
	\bibitem [{\citenamefont {{ASTM E1621
				22}}(2022)}]{ASTME1621222022StandardSpectrometry}%
	\BibitemOpen
	\bibfield  {author} {\bibinfo {author} {\bibnamefont {{ASTM E1621 22}}},\
	}\href {\doibase 10.1520/E1621-22} {\enquote {\bibinfo {title} {{Standard
					Guide for Elemental Analysis by Wavelength Dispersive X-Ray Fluorescence
					Spectrometry}},}\ } (\bibinfo {year} {2022})\BibitemShut {NoStop}%
	\bibitem [{\citenamefont {{ASTM International G99
				23}}(2023)}]{ASTMInternationalG99232023TestApparatus}%
	\BibitemOpen
	\bibfield  {author} {\bibinfo {author} {\bibnamefont {{ASTM International G99
					23}}},\ }\href {\doibase 10.1520/G0099-23} {\emph {\bibinfo {title} {{Test
					Method for Wear Testing with a Pin-on-Disk Apparatus}}}},\ \bibinfo {type}
	{Tech. Rep.}\ (\bibinfo  {institution} {ASTM International},\ \bibinfo
	{address} {West Conshohocken, PA},\ \bibinfo {year} {2023})\BibitemShut
	{NoStop}%
	\bibitem [{\citenamefont {{ASTM G133-22}}(2022)}]{ASTMG133-222022TestWear}%
	\BibitemOpen
	\bibfield  {author} {\bibinfo {author} {\bibnamefont {{ASTM G133-22}}},\
	}\href {\doibase 10.1520/G0133-22} {\emph {\bibinfo {title} {{Test Method for
					Linearly Reciprocating Ball-on-Flat Sliding Wear}}}},\ \bibinfo {type} {Tech.
		Rep.}\ (\bibinfo  {institution} {ASTM International},\ \bibinfo {address}
	{West Conshohocken, PA},\ \bibinfo {year} {2022})\BibitemShut {NoStop}%
	\bibitem [{\citenamefont {Kumar}\ \emph {et~al.}(2026)\citenamefont {Kumar},
		\citenamefont {Mukherjee},\ and\ \citenamefont
		{Singh}}]{Kumar2026SynergisticTemperature}%
	\BibitemOpen
	\bibfield  {author} {\bibinfo {author} {\bibfnamefont {A.}~\bibnamefont
			{Kumar}}, \bibinfo {author} {\bibfnamefont {R.}~\bibnamefont {Mukherjee}}, \
		and\ \bibinfo {author} {\bibfnamefont {M.~K.}\ \bibnamefont {Singh}},\ }\href
	{\doibase 10.1016/j.wear.2025.206384} {\bibfield  {journal} {\bibinfo
			{journal} {Wear}\ }\textbf {\bibinfo {volume} {584-585}},\ \bibinfo {pages}
		{206384} (\bibinfo {year} {2026})}\BibitemShut {NoStop}%
	\bibitem [{\citenamefont {Kumar}\ \emph
		{et~al.}(2024{\natexlab{b}})\citenamefont {Kumar}, \citenamefont {Rachwani},\
		and\ \citenamefont {Kumar~Singh}}]{Kumar2024DrySubstrates}%
	\BibitemOpen
	\bibfield  {author} {\bibinfo {author} {\bibfnamefont {A.}~\bibnamefont
			{Kumar}}, \bibinfo {author} {\bibfnamefont {H.}~\bibnamefont {Rachwani}}, \
		and\ \bibinfo {author} {\bibfnamefont {M.}~\bibnamefont {Kumar~Singh}},\
	}\href {\doibase 10.1177/13506501241227478} {\bibfield  {journal} {\bibinfo
			{journal} {Proceedings of the Institution of Mechanical Engineers, Part J:
				Journal of Engineering Tribology}\ }\textbf {\bibinfo {volume} {238}},\
		\bibinfo {pages} {662} (\bibinfo {year} {2024}{\natexlab{b}})}\BibitemShut
	{NoStop}%
\end{thebibliography}
%\bibliographystyle{apsrev4-1}


%

\end{document}